\pdfoutput=1
\documentclass[preprint,showpacs,preprintnumbers,amsmath,amssymb,nofootinbib]{revtex4}

\usepackage[dvips]{graphicx}% Include figure files

\usepackage{dcolumn}% Align table columns on decimal point
\usepackage{bm}% bold math
\usepackage{color}

\usepackage{bibentry}
\usepackage{url}

\begin{document}

%\preprint{...}

\title{What can(not) be measured with ton-scale dark matter direct detection experiments}

\author{Miguel Pato}
\email{migpato@physik.uzh.ch}
\affiliation{Institute for Theoretical Physics, University of Z\"urich, Winterthurerstrasse 190, CH-8057, Switzerland}
%\affiliation{Institut d'Astrophysique de Paris, UMR 7095-CNRS, Univ.~Pierre \& Marie Curie, 98bis Bd Arago 75014 Paris, France}
%\affiliation{Dipartimento di Fisica, Universit\`a degli Studi di Padova, via Marzolo 8, I-35131, Padova, Italy}

\date{\today}% It is always \today, today,
             %  but any date may be explicitly specified

\begin{abstract}
Direct searches for dark matter have prompted in recent years a great deal of excitement within the astroparticle physics community, but the compatibility between signal claims and null results of different experiments is far from being a settled issue. In this context, we study here the prospects for constraining the dark matter parameter space with the next generation of ton-scale detectors. Using realistic experimental capabilities for a wide range of targets (including fluorine, sodium, argon, germanium, iodine and xenon), the role of target complementarity is analysed in detail while including the impact of astrophysical uncertainties in a self-consistent manner. We show explicitly that a multi-target signal in future direct detection facilities can determine the sign of the ratio of scalar couplings $f_n/f_p$, but not its scale. This implies that the scalar-proton cross-section is left essentially unconstrained if the assumption $f_p\sim f_n$ is relaxed. Instead, we find that both the axial-proton cross-section and the ratio of axial couplings $a_n/a_p$ can be measured with fair accuracy if multi-ton instruments using sodium and iodine will eventually come online. Moreover, it turns out that future direct detection data can easily discriminate between elastic and inelastic scatterings. Finally, we argue that, with weak assumptions regarding the WIMP couplings and the astrophysics, only the dark matter mass and the inelastic parameter (i.e.~mass splitting) may be inferred from the recoil spectra -- specifically, we anticipate an accuracy of tens of GeV (tens of keV) in the measurement of the dark matter mass (inelastic parameter).

%Similarly, the discrimination between scalar and vector couplings appears virtually impossible in the near future.

\end{abstract}

% PACS, the Physics and Astronomy
%                             % Classification Scheme.
%\keywords{Suggested keywords}%Use showkeys class option if keyword
                              %display desired
\maketitle

\section{Introduction}\label{secintro}

\par Although extensive evidence supports the presence of large quantities of dark matter (DM) in the Universe, its specific nature remains undisclosed. Over the last decades, different, complementary techniques have been put forward in the effort to identify dark matter and constrain its properties. In particular, direct detection appears very promising in the near future and lies today at the cutting edge of research in particle physics and astrophysics. The idea behind direct detection is rather simple and compelling: if the dark matter in our Galaxy is composed of Weakly Interacting Massive Particles (WIMPs), then the Earth is continuously crossing a large flux of WIMPs, which can be detected by their scattering off nuclei in underground instruments. The past few years have seen a great excitement in direct dark matter searches, mainly due to a mix of both startling signals and null results. On the one hand, the DAMA collaboration reports an annual modulation of WIMP-like events with impressive statistical significance \cite{DAMA2008,DAMA2010}, whereas CoGeNT detects excess events towards the threshold energy \cite{cogent} also with an annual modulation \cite{cogentannualmod}. Under standard assumptions the two results are not compatible, but it is worth noting that there is enough freedom on the particle physics, astrophysics and experimental sides in order to explain both results with the same WIMP candidate (see e.g.~\cite{HooperCoGeNT,ChangIso,FengIso,Arina2011,FrandsenIso,DelNobile:2011je}). On the other hand, there is a whole array of direct detection experiments reporting no events above the expected background. Among them, XENON10/100 \cite{Xenon10,Xenon100,Xenon1002} and CDMS \cite{CDMS2009,cdms10,CDMSEDEL2011} are perhaps the most sensitive and report small numbers of WIMP-like scatterings largely compatible with the background. Also CRESST \cite{cresst2011} and EDELWEISS \cite{EDEL2011} see events in their signal region, but an interpretation of the latter in terms of WIMP scattering appears rather difficult. A way out of this puzzling situation may be provided by the upcoming generation of direct detection experiments which feature ton-scale or multi-ton-scale targets. Hopefully, these instruments will shed light on the topic over the next decade and confirm or rule out the dark matter hypothesis.

\par Recently, several works have appeared that discuss the prospects for measuring different WIMP-nucleon couplings with upcoming direct detection data. For instance, in Ref.~\cite{Vergados} (see also \cite{Bednyakov1994,Bednyakov2008}) the authors address the possibility of extracting the scalar-proton, axial-proton and axial-neutron cross-sections with the total event rates in future fluorine, germanium and iodine targets. In Refs.~\cite{Shan1,Shan2} (see also \cite{Drees2007,Drees2008}) a different method is presented to estimate scalar and axial couplings as well as the ratios of proton to neutron couplings. These and other works rely on a certain degree of theoretical assumptions regarding the WIMP-nucleon couplings -- e.g.~one usually assumes that the scalar cross-section dominates over the axial cross-section (or vice-versa) or that scalar-proton and scalar-neutron couplings are equal. Although these are valid assumptions for the most studied WIMP candidates such as the supersymmetric neutralino (see \cite{EllisDD}), we should keep in mind that other possibilities exist and lead to a rich phenomenology. For example, it is possible to construct models where the WIMP scatters off protons and neutrons with different amplitudes -- specific models have been presented in \cite{FengIso,ChangIso,KangIso}. This implies that spin-independent rates no longer scale as $A^2$ ($A$ being the target mass number) as usually assumed. Also, as pointed out in the literature recently \cite{Green2007,Green1,Drees2007,Drees2008,StrigariTrotta,fox1,fox2,Akrami1,Akrami2,paperastrodirect,APeter,APeter2011} but also ten years ago \cite{Fornengoastro1,BelliFornengo}, the unknown WIMP local density and velocity distribution translate into a sizeable ``astrophysical'' uncertainty that hinders the determination of WIMP properties and has not been addressed in works such as \cite{Vergados}\footnote{Let us stress that other methods \cite{Drees2008,Shan1,Shan2,fox1,fox2} have also been developed to extract WIMP properties independently from astrophysical unknowns.}. In this context, the aim of the present paper is to identify what kind of information can really be measured in a robust manner with the next generation of ton-scale direct detection experiments. We focus our attention on the extraction of WIMP mass, couplings (scalar, vector, axial) and inelastic parameter with special emphasis on the importance of target complementarity. Ours is an analysis that uses realistic upcoming experimental capabilities and keeps particle physics theoretical assumptions to a minimum. We show explicitly the effect of relaxing certain widely used assumptions (e.g.~the equality of scalar-proton and scalar-neutron couplings) on the constrained WIMP parameter space; this issue has been ignored in almost all past studies. Moreover, the astrophysical uncertainty is folded self-consistently in our results and its role is discussed in detail. We perform a Bayesian analysis which allows the translation of statistical and astrophysical uncertainties into the derived constraints, thus improving upon analytical approaches as the one presented in \cite{Vergados}. Another point addressed here is the possibility of distinguishing elastic and inelastic WIMPs and measuring the corresponding inelastic parameter. This study assumes particular relevance now since complementarity analyses including searches at the Large Hadron Collider will soon be possible, and hence it is essential to know what can -- and, perhaps more importantly, what cannot -- be learnt from direct detection data alone.

\par The paper is organised as follows. In Section \ref{secform} we briefly review the formalism of direct searches before specifying the experimental capabilities and dark matter benchmarks in Section \ref{secexp}. Section \ref{secmeth} outlines our Bayesian methodology, while in Section \ref{secres} we present the results. Finally, the main conclusions are drawn in Section \ref{secconc}.

\section{Direct detection formalism}\label{secform}

\par The central quantity in direct detection studies is the differential rate $dR/dE_R$ at which WIMP-nucleus scattering events occur in an underground instrument composed of nuclei $N(A,Z)$ (for reviews see \cite{Jungman,LewinSmith,BertoneBook}). This rate is simply given by the convolution of the WIMP flux as seen from the Earth and the scattering cross-section, and it can be conveniently written as 
\begin{equation}\label{dRdE}
\frac{dR}{dE_R} (E_R) = \frac{\rho_0}{2 m_\chi \mu_N^2} \times \left[ \sigma_{\chi-N}^{SD,0} F_{SD}^2(E_R) + \sigma_{\chi-N}^{SI,0} F_{SI}^2(E_R) \right] \times \mathcal{F}(v_{min}(E_R))
\end{equation}
in units of counts/ton/yr/keV and where $\rho_0$ is the local WIMP mass density, $m_\chi$ is the WIMP mass, $\mu_N$ is the WIMP-nucleus reduced mass and $\sigma_{\chi-N}^{SD,0}$ ($\sigma_{\chi-N}^{SI,0}$) is the zero-momentum spin-dependent (spin-independent) cross-section. The form factors $F_{SD}$ and $F_{SI}$ account for the energy dependence of the respective cross-sections. Finally, the factor $\mathcal{F}$ is the WIMP mean inverse velocity and reads
\begin{equation}\label{fastro}
\mathcal{F}(v_{min}) = \int_{v>v_{min}}{d^3\vec{v} \, \frac{f(\vec{v}+\vec{v}_e)}{v}} \quad ,
\end{equation}
$f$ being the local WIMP velocity distribution and $\vec{v}_e$ the Earth velocity in the galactic rest frame. The minimum WIMP velocity that produces a nuclear recoil of energy $E_R$ is 
\begin{equation}\label{vmin}
v_{min}(E_R) = \frac{1}{\sqrt{2m_N E_R}} \left(\frac{m_N E_R}{\mu_N} +\delta\right) \quad ,
\end{equation}
where $m_N$ is the nucleus mass and $\delta$ is the inelastic parameter that vanishes in the case of elastic scattering and reads $\delta=m_{\chi'}-m_{\chi}$ in the case of inelastic dark matter models ($\chi'$ being the excited DM state). 

\par In equation \eqref{dRdE}, the astrophysical dependence is encoded in $\rho_0$ and $\mathcal{F}$, while the particle and nuclear physics enter in the middle factor. Regarding astrophysics, we follow the approach of Refs.~\cite{Lisanti,paperastrodirect} and model the WIMP velocity distribution as
\begin{equation}\label{fv}
f(w) = \left\{
\begin{array}{ll}
N_f \left(\exp\left(\frac{v_{esc}^2-w^2}{k v_0^2}\right)-1 \right)^k & \textrm{for }w<v_{esc} \\
0 & \textrm{for }w\geq v_{esc} 
\end{array}
   \right. \quad ,
\end{equation}
where $N_f$ is a suitable normalisation, $k$ is a shape parameter, $v_{esc}$ is the local escape velocity and $v_0$ traces the velocity dispersion. Different parameterisations for the WIMP phase space distribution (including dark disks and streams) and the corresponding effect in direct detection are studied in \cite{Green1,APeter2011}. Also, since we are not interested in modelling the annual modulation signal, the Earth orbit and the Sun's peculiar motion are disregarded so that $\vec{v}_e$ is given by the local circular velocity $\vec{v}_c^0$, whose absolute value we shall identify with $v_0$ (see Ref.~\cite{paperastrodirect} for a detailed discussion).

\par The WIMP-nucleon scattering cross-section is commonly split into two components, spin-dependent (SD) and spin-independent (SI). The former arises from the WIMP-quark axial coupling and is expressed as 
\begin{equation}\label{sigmaSD}
\sigma_{\chi-N}^{SD,0} = \frac{32}{\pi} \mu_N^2 G_F^2 \frac{J_N+1}{J_N} \left( a_p \langle S_p^N\rangle + a_n \langle S_n^N\rangle \right)^2 \quad ,
\end{equation}
in which $G_F$ is the Fermi coupling constant, $J_N$ is the spin of the target nucleus $N$, $a_p$ ($a_n$) is the axial WIMP-proton (-neutron) coupling and $\langle S_p^N\rangle$ ($\langle S_n^N\rangle$) is the expectation value of the spin of protons (neutrons) in the nucleus $N$. The values of $\langle S_p^N\rangle$ and $\langle S_n^N\rangle$ for different nuclei may be found in Ref.~\cite{BednyakovSpin}. Defining the axial WIMP-proton cross-section as $\sigma_p^{SD}\equiv \frac{32}{\pi} \mu_p^2 G_F^2 \frac{3}{4} a_p^2$ (recall that $J_p=1/2$, $\langle S_p^p\rangle=1/2$, $\langle S_n^p\rangle=0$), we recast equation \eqref{sigmaSD} as 
\begin{equation}\label{sigmaSD2}
\sigma_{\chi-N}^{SD,0} = \frac{4 \mu_N^2}{3 \mu_p^2}  \frac{J_N+1}{J_N} \sigma_p^{SD} \left( \langle S_p^N\rangle + \langle S_n^N\rangle a_n/a_p \right)^2 \quad ,
\end{equation}
where $\mu_p$ is the WIMP-proton reduced mass. In the body of the paper, we shall use $\sigma_p^{SD}$ and $a_n/a_p$ as our independent phenomenological parameters. The spin-dependent nuclear form factor $F_{SD}$ is implemented according to the standard prescription \cite{Jungman}:
\begin{equation}\label{FSD}
F_{SD}^2(E_R) = \frac{S(q)}{S(0)}
\end{equation}
with $q^2=2m_N E_R$ and 
\begin{equation}
S(q) = a_0^2 S_{00}(q) + a_1^2 S_{11}(q) + a_0 a_1 S_{01}(q) \quad ,
\end{equation}
where $a_0=a_p+a_n$ and $a_1=a_p-a_n$. The structure functions $S_{00}$, $S_{11}$ and $S_{01}$ depend on the target nuclei and are parameterised following Ref.~\cite{BednyakovFormFactor} for the elements used in this work (cf.~next Section and Table \ref{tabSii}).

\par Spin-independent scattering arises, instead, from the WIMP-quark scalar and vector couplings. The corresponding cross-sections are usually written in terms of the scalar and vector WIMP-nucleon couplings $f_{p,n}$ and $b_{p,n}$ respectively:
\begin{equation}\label{sigmaSI}
\sigma_{\chi-N}^{SI,0} = \frac{4}{\pi} \mu_N^2 \left[ \left(Z f_p + (A-Z)f_n \right)^2 + \frac{1}{256} \left(Z b_p + (A-Z)b_n\right)^2   \right] \quad . 
\end{equation}
Similarly to the SD case, we define the scalar WIMP-proton cross-section $\sigma_p^{sc}\equiv \frac{4}{\pi}\mu_p^2 f_p^2$ and the vector WIMP-neutron cross-section $\sigma_n^{vec}\equiv \frac{1}{64\pi}\mu_n^2 b_n^2$ (the reason for defining $\sigma_n^{vec}$ and not $\sigma_p^{vec}$ will become apparent in Sections \ref{secexp} and \ref{secSI}) to recast equation \eqref{sigmaSI} as
\begin{equation}\label{sigmaSI2}
\sigma_{\chi-N}^{SI,0} = \frac{\mu_N^2}{\mu_p^2} \left[ \sigma_p^{sc} \left(Z + (A-Z)f_n / f_p \right)^2 + \sigma_n^{vec} \left(Z b_p /b_n + (A-Z)\right)^2   \right] \quad ,
\end{equation}
where we made the very good approximation $\mu_n\simeq \mu_p$. Again, we will be using the cross-sections $\sigma_p^{sc}$, $\sigma_n^{vec}$ and the coupling ratios $f_n/f_p$, $b_p/b_n$ as our phenomenological parameters. As for the spin-independent nuclear form factor $F_{SI}$, we take the parameterisation introduced in \cite{LewinSmith} and that has been shown to be a reliable approximation at least for $E_R < 100$ keV \cite{GondoloFormFactor}:
\begin{equation}
F_{SI}(E_R)=3\frac{\textrm{sin}(q r_n)-(q r_n)\textrm{cos}(q r_n)}{(q r_n)^3} \textrm{exp}(-(q s)^2/2) \quad ,
\end{equation}
where $q$, $r_n$ and $s$ are implicitly expressed in natural units, $r_n^2=c^2+\frac{7}{3}\pi^2 a^2-5 s^2$, $s\simeq 0.9$ fm, $a\simeq0.52$ fm and $c=(1.23A^{1/3}-0.6)\textrm{fm}$.

\par Now, once one defines a WIMP model ($m_\chi$, $\sigma_p^{SD}$, $a_n/a_p$, $\sigma_p^{sc}$, $f_n/f_p$, $\sigma_n^{vec}$, $b_p/b_n$, $\delta$), an astrophysical setup ($\rho_0$, $v_0$, $v_{esc}$, $k$) and a target nucleus $N$, the differential event rate $dR/dE_R$ in equation \eqref{dRdE} is unambiguously fixed. But what direct detection experiments can actually observe is the number of events in a given energy bin $\left[E_1,E_2\right]$:
\begin{equation}\label{NE1E2}
N_R(E_1,E_2) =\epsilon_{eff} \int_{E_1}^{E_2}{ dE_R \, \frac{d\tilde{R}}{dE_R}(E_R) }
\end{equation}
with 
\begin{equation}\label{dRdEtilde}
\frac{d\tilde{R}}{dE_R}(E_R) = \int_0^\infty{dE'_R \, \frac{1}{\sqrt{2\pi}\sigma(E'_R)} \exp\left(-\frac{(E_R-E'_R)^2}{2\sigma^2(E'_R)}\right) \frac{dR}{dE_R}(E'_R)  } \quad ,
\end{equation}
where $\epsilon_{eff}$ is the mean effective exposure of the experiment and $\sigma$ its energy resolution. The number of counts in equation \eqref{NE1E2} plays a central role in the remaining of this work.

\section{Experimental capabilities \& dark matter benchmarks}\label{secexp}

\par We are interested in assessing the prospects for direct detection experiments over the next decade. Hence we focus on a wide range of target elements that are being pushed forward by several collaborations, namely $^{19}$F, $^{23}$Na, $^{40}$Ar, $^{73}$Ge, $^{127}$I and $^{131}$Xe. The respective properties and experimental figures are outlined in Table \ref{tabExp}. In view of the developments in the field of direct detection, we consider ton-scale and multi-ton-scale instruments with $\sim \mathcal{O}(10)$ keV threshold energies and energy resolutions parameterised as
\begin{equation}
\sigma(E_R) = \left( a + b\sqrt{E_R/\textrm{keV}} + c\, E_R/\textrm{keV} \right)^d \textrm{ keV} \quad .
\end{equation}

\begin{table*}[htp]
\centering
\fontsize{11}{11}\selectfont
\begin{tabular}{c||c}
\hline
\hline

Element & $S_{00}$, $S_{11}$, $S_{01}$ \\
\hline
$^{19}$F   &  eq.~(7) of Ref.~\cite{BednyakovFormFactor} \\   
$^{23}$Na  &  eq.~(10) of Ref.~\cite{BednyakovFormFactor} \\     
$^{73}$Ge  &  eq.~(17) of Ref.~\cite{BednyakovFormFactor} \\    
$^{127}$I  &  eq.~(20) of Ref.~\cite{BednyakovFormFactor}, ``Bonn A'' coefficients from Table VI  \\   
$^{131}$Xe &  eq.~(21) of Ref.~\cite{BednyakovFormFactor}, ``Bonn A'' coefficients from Table IX  \\ 
\hline
\end{tabular}
\caption{\fontsize{9}{9}\selectfont The parameterisation of the SD structure functions $S_{00}$, $S_{11}$ and $S_{01}$ used in this work.}\label{tabSii}
\end{table*}

\begin{table*}[htp]
\centering
\fontsize{11}{11}\selectfont
\begin{tabular}{c||ccccccc||cccccc}
\hline
\hline

Element & $A$ & $Z$ & $A-Z$ & $\langle S_p^N\rangle$ & $\langle S_n^N\rangle$ & $J_N$ & $m_N\,[m_u]$ & $\epsilon_{eff}\,[\textrm{ton.yr}]$ & $E_{thr}\,[\textrm{keV}]$ & $a$ & $b$ & $c$ & $d$ \\
\hline
$^{19}$F   &  19 & 9 & 10    & 0.4751 & $-$0.0087     & 1/2 & 18.998 & 1.00 & 10 & 5.0 & 0.0 & 0.0 & 1.0 \\   
$^{23}$Na  &  23 & 11 & 12   & 0.2477 & 0.0199      & 3/2 & 22.990 & 1.00 & 10 & 0.0 & 0.245 & 0.0027 & 1.0 \\  
$^{40}$Ar  &  40 & 18 & 22   & 0      & 0           & --   & 39.962 & 6.40 & 30 & 0.0 & 0.7 &0.0 & 1.0 \\  
$^{73}$Ge  &  73 & 32 & 41   & 0.030  & 0.378       & 9/2 & 72.923 & 2.16 & 10 & 0.3$^2$ & 0.0 & 0.06$^2$ & 0.5 \\  
$^{127}$I  &  127& 53 & 74   & 0.309  & 0.075       & 5/2 & 126.904& 1.00 & 10 & 0.0 &  0.134 & 0.0008 & 1.0 \\ 
$^{131}$Xe &  131& 54 & 77   & $-$0.041 & $-$0.236      & 3/2 & 130.905& 2.00 & 10 & 0.0 & 0.6 & 0.0 & 1.0 \\

\hline
\end{tabular}
\caption{\fontsize{9}{9}\selectfont The nuclear properties of various target elements and the corresponding experimental capabilities. The values for $\langle S_p^N\rangle$ and $\langle S_n^N\rangle$ are taken from Ref.~\cite{BednyakovSpin} and the nuclei mass $m_N$ is presented in atomic mass units $m_u=0.931494\textrm{ GeV/c}^2$. In Section \ref{secSISD} we shall use an exposure for $^{23}$Na and $^{127}$I of 3 ton.yr instead of the baseline value of 1 ton.yr shown here. See the text for further details.}\label{tabExp}
\end{table*}

\par Today, two techniques look particularly promising to assemble low-background, ton-scale instruments: cryogenic detectors at mK temperatures (using for instance $^{73}$Ge as target material) and noble liquid detectors (featuring for instance $^{40}$Ar or $^{131}$Xe as targets). The former technique is being developed and improved by EURECA \cite{eureca} and SuperCDMS/GEODM \cite{geodm}, whereas the latter will be used in DARWIN \cite{darwin}, MAX \cite{max} and XMASS \cite{xmass}, as detailed in Ref.~\cite{paperastrodirect} to which we refer for further discussion. We assume the same experimental capabilities for $^{40}$Ar, $^{73}$Ge and $^{131}$Xe as in that paper\footnote{For the energy resolutions, see in particular Ref.~\cite{Xeresolution} for Xe and Ref.~\cite{Ahmed:2009rh} for Ge.} and summarise the relevant figures in Table \ref{tabExp}.

\par Another interesting target, particularly due its SD sensitivity, is $^{19}$F, already at use in experiments as COUPP \cite{COUPP2008,COUPP2010}. We take an optimistic (but presumably realistic) exposure of $\epsilon_{eff}(^{19}\textrm{F})=1$ ton.yr and a threshold energy $E_{thr}(^{19}\textrm{F})=10$ keV. Although COUPP cannot in principle yield spectral information, we assume a constant energy resolution of $\sigma(^{19}\textrm{F})=5$ keV, in line with that obtained in NaF bolometers \cite{NaFbol}.

\par At last, in order to fully explore target complementarity, also $^{23}$Na and $^{127}$I will be considered. These are elements employed in different instruments including DAMA/LIBRA \cite{DAMA2008,DAMA2010}; iodine is also present in the CF$_3$I target material of COUPP \cite{COUPP2008,COUPP2010}. Similarly to fluorine, we consider baseline exposures $\epsilon_{eff}(^{23}\textrm{Na})=\epsilon_{eff}(^{127}\textrm{I})=1$ ton.yr and threshold energies $E_{thr}(^{23}\textrm{Na})=E_{thr}(^{127}\textrm{I})=10$ keV. For both materials we take a DAMA-like resolution \cite{DAMA2008b}, $\sigma(X=\textrm{Na},\textrm{I})=(0.448\sqrt{q_X E_R/\textrm{keV}} + 0.0091\, q_X E_R/\textrm{keV})\textrm{ keV}$ (where $q_X$ is the quenching factor of target X; $q_{Na}=0.3$ and $q_I=0.09$).

\par For all targets presented in Table \ref{tabExp}, we set a maximum recoil energy of 100 keV. It is worth noticing that higher energies can lead to tighter WIMP constraints as studied in \cite{APeter2011}, but it is still not clear if the usual nuclear form factors are valid and if experimental background is an issue in that energy range.

\vspace{0.5cm}

\par The last ingredient to specify before proceeding with the analysis is the set of DM benchmarks to study. We take the WIMP models shown in Table \ref{tabBench} that feature various coupling configurations. In particular, the SD and SI cross-sections are fixed to values below the latest upper limits, $10^{-5}$ pb \cite{XENONSD} and $10^{-9}$ pb \cite{Xenon1002} respectively. The reason why we focus on 50 GeV WIMPs is because most direct detection experiments have an optimal sensitivity around this mass -- we are indeed interested in extracting the maximum possible information from future data and, in that respect, ours is an optimistic work. However, WIMP masses of 25 and 250 GeV will also be extensively used. In order to determine the accuracy in the measurement of the inelastic parameter $\delta$, we shall also consider a DM benchmark with $\delta=40$ keV in the ballpark of the values explored in the literature \cite{InelasticWeiner}, and $\delta=100$ keV in light of recent exclusion limits \cite{Angle:2009xb,Ahmed:2010hw,Xenon100Ine}. Table \ref{tabNevts} shows the expected total event number for the various DM benchmarks in Table \ref{tabBench} and targets in Table \ref{tabExp} considering a recoil energy range between the threshold energy and 100 keV.

\begin{table*}[htp]
\centering
\fontsize{11}{11}\selectfont
\begin{tabular}{c||cccccccc}
\hline
\hline
DM benchmark & $m_\chi\,[\textrm{GeV}]$ & $\sigma_p^{SD}\,[\textrm{pb}]$ & $a_n/a_p$ & $\sigma_p^{sc}\,[\textrm{pb}]$ & $f_n/f_p$ & $\sigma_n^{vec}\,[\textrm{pb}]$ & $b_p/b_n$ & $\delta\,[\textrm{keV}]$ \\
\hline
1 & 50 & 0 & $-$1.0 & $10^{-9}$ & 1.0 & 0 & 0.0 & 0 \\
1a & 25 & 0 & $-$1.0 & $10^{-9}$ & 1.0 & 0 & 0.0 & 0 \\
1b & 250 & 0 & $-$1.0 & $10^{-9}$ & 1.0 & 0 & 0.0 & 0 \\
2 & 50 & 0 & $-$1.0 & $10^{-9}$ & 1.0 & $10^{-9}$ & 0.0 & 0 \\
3 & 50 & $10^{-5}$ & $-$1.0 & $10^{-9}$ & 1.0 & 0 & 0.0 & 0 \\
3a & 50 & $10^{-3}$ & $-$1.0 & $10^{-9}$ & 1.0 & 0 & 0.0 & 0 \\
3b & 50 & $10^{-5}$ & 0.0 & $10^{-9}$ & 1.0 & 0 & 0.0 & 0 \\
3c & 50 & $10^{-5}$ & $+$1.0 & $10^{-9}$ & 1.0 & 0 & 0.0 & 0 \\
3d & 50 & $10^{-3}$ & 0.0 & $10^{-9}$ & 1.0 & 0 & 0.0 & 0 \\
3e & 50 & $10^{-3}$ & $+$1.0 & $10^{-9}$ & 1.0 & 0 & 0.0 & 0 \\
3f & 25 & $10^{-5}$ & $-$1.0 & $10^{-9}$ & 1.0 & 0 & 0.0 & 0 \\
3g & 250 & $10^{-5}$ & $-$1.0 & $10^{-9}$ & 1.0 & 0 & 0.0 & 0 \\
3h & 25 & $10^{-3}$ & $-$1.0 & $10^{-9}$ & 1.0 & 0 & 0.0 & 0 \\
3i & 250 & $10^{-3}$ & $-$1.0 & $10^{-9}$ & 1.0 & 0 & 0.0 & 0 \\
4 & 50 & 0 & $-$1.0 & $10^{-9}$ & 1.0 & 0 & 0.0 & 40 \\
4a & 50 & 0 & $-$1.0 & $10^{-5}$ & 1.0 & 0 & 0.0 & 100 \\
4b & 25 & 0 & $-$1.0 & $10^{-9}$ & 1.0 & 0 & 0.0 & 40 \\
4c & 250 & 0 & $-$1.0 & $10^{-9}$ & 1.0 & 0 & 0.0 & 40 \\
5 & 50 & $10^{-5}$ & $-$1.0 & $10^{-9}$ & 1.0 & 0 & 0.0 & 40 \\
5a & 25 & $10^{-5}$ & $-$1.0 & $10^{-9}$ & 1.0 & 0 & 0.0 & 40 \\
5b & 250 & $10^{-5}$ & $-$1.0 & $10^{-9}$ & 1.0 & 0 & 0.0 & 40 \\

\hline
\end{tabular}
\caption{\fontsize{9}{9}\selectfont The properties of the DM benchmarks used to assess the prospects of the upcoming generation of ton-scale direct detection experiments.}\label{tabBench}
\end{table*}

\begin{table*}[htp]
\centering
\fontsize{11}{11}\selectfont
\begin{tabular}{c||cccccccc}
\hline
\hline
DM benchmark & $^{19}$F & $^{23}$Na  & $^{40}$Ar  & $^{73}$Ge  & $^{127}$I  & $^{131}$Xe \\
\hline
1 & 11  & 18  & 125 & 303 & 183 & 362 \\
1a& 10  & 16  & 28  & 161 & 71  & 142 \\
1b& 3   & 6   & 88  & 145 & 70  & 159 \\
2 & 14  & 21  & 163 & 399 & 245 & 488 \\
3 & 282 & 53  & 125 & 387 & 193 & 378 \\
3a(*)& 26975 & 11044 & 125 & 8761 & 3548 & 1955 \\
3b(*)& 271 & 180 & 125 & 305 & 603 & 371 \\ 
3c(*)& 262 & 203 & 125 & 434 & 634 & 406 \\
3d(*)& 26114 & 13046 & 125 & 454 & 5863 & 1053 \\
3e(*)& 25168 & 15216 & 125 & 13330 & 8887 & 4549 \\
3f(*)& 284 & 141 & 28 & 211 & 224 & 148 \\
3g(*)& 101 & 61 & 88 & 176 & 260 & 168 \\
3h(*)& 27303 & 9691 & 28 & 5113 & 1374 & 758 \\
3i(*)& 9594  & 4251 & 88 & 3626 & 1731 & 977 \\
4 & $<$1   & $<$1   & 4   & 28  & 31  & 64  \\
4a& $<$1   & $<$1   & $<$1   & $<$1   & 425 & 1025 \\
4b& $<$1   & $<$1   & $<$1   & 2   & 2   & 4 \\
4c& $<$1   & $<$1   & 14  & 40  & 38  & 72 \\
5 & $<$1   & $<$1   & 4   & 38  & 33  & 66  \\
5a& $<$1   & $<$1   & $<$1   & 3   & 2   & 4  \\
5b& $<$1   & $<$1   & 14  & 52  & 37  & 75 \\
\hline
\end{tabular}
\caption{\fontsize{9}{9}\selectfont The expected total number of events in each of the targets in study and for the various DM benchmarks. The recoil energy range considered spans from the threshold energy up to 100 keV. The experimental capabilities assumed are the ones in Table \ref{tabExp} except for the lines marked with (*) in which an exposure of 3 ton.yr was taken for $^{23}$Na and $^{127}$I. Notice that even in the cases where a very small number of events ($<$1) is expected, the corresponding target data is valuable in constraining the WIMP parameter space.}\label{tabNevts}
\end{table*}

\section{Methodology}\label{secmeth}

\par Given the formalism presented in Section \ref{secform} and the DM benchmarks and experimental capabilities specified in Section \ref{secexp}, we start by generating mock (``true'') data for each benchmark and each target nucleus assuming our fiducial astrophysical model: $\rho_0=0.4\textrm{ GeV/cm}^3$, $v_0=230$ km/s, $v_{esc}=544$ km/s, $k=1$. Ten linearly-spaced energy bins are constructed between the relevant threshold energy and 100 keV. For the different DM benchmarks and target materials, bin counts range from a few to several hundreds. Along the work, we shall use the mock data corresponding to a specific target, or a combination of data sets as shown in Table \ref{tabData}. Data I employ nuclei widely spread across atomic number $Z$; data II feature the targets used in Ref.~\cite{Vergados}; data III include all targets and, finally, data IV are a variant of data III with enhanced $^{23}$Na and $^{127}$I exposures (3 ton.yr). The usefulness of each data set will become apparent in the next Section.

\begin{table*}[htp]
\centering
\fontsize{11}{11}\selectfont
\begin{tabular}{l||llll}
\hline
\hline
Data set & targets & & & comments \\
\hline
data I   & $^{19}$F+$^{40}$Ar+$^{73}$Ge+$^{131}$Xe & & & -- \\
data II  & $^{19}$F+$^{73}$Ge+$^{127}$I & & & --\\
data III & $^{19}$F+$^{23}$Na+$^{40}$Ar+$^{73}$Ge+$^{127}$I+$^{131}$Xe & & &-- \\
data IV  & $^{19}$F+$^{23}$Na+$^{40}$Ar+$^{73}$Ge+$^{127}$I+$^{131}$Xe & & & $\epsilon_{eff}(^{23}\textrm{Na})=\epsilon_{eff}(^{127}\textrm{I})=3$ ton.yr\\

\hline
\end{tabular}
\caption{\fontsize{9}{9}\selectfont The mock experimental data sets used along the work.}\label{tabData}
\end{table*}

\par The next step is to scan over the parameter space composed by the WIMP properties ($m_\chi$, $\sigma_p^{SD}$, $a_n/a_p$, $\sigma_p^{sc}$, $f_n/f_p$, $\sigma_n^{vec}$, $b_p/b_n$, $\delta$) and the astrophysical unknowns ($\rho_0$, $v_0$, $v_{esc}$, $k$), whose ranges are outlined in Table \ref{tabPars}. To this end, we use the MultiNest code \cite{MultiNest1,MultiNest2,MultiNest3}\footnote{The author acknowledges Louis Strigari and Roberto Trotta for kindly providing access to their private direct detection code.} which is a very efficient sampler of higher-dimensionality parameter spaces and relies on Bayes' theorem to update the prior $p(\theta)$:
\begin{equation}
p(\theta|d) = \frac{\mathcal{L}(\theta)p(\theta)}{p(d)} \quad ,
\end{equation}
where $p(\theta|d)$ is the posterior probability function, $\theta$ ($d$) represents the parameter (data) set, $\mathcal{L}(\theta)$ is the likelihood of the parameter set $\theta$ given the data $d$ and $p(d)$ is the so-called Bayesian evidence (for a review on Bayesian tools and techniques see Ref.~\cite{TrottaReview}). In this work the MultiNest code is run with \url{nlive=3000} live points, an efficiency parameter \url{eff=0.3} and a tolerance \url{tol=0.8} -- these parameters were found to be appropriate to scan efficiently and in a reasonable period of time the multi-dimensional parameter space in study; see \cite{superbayes} for further details. Since we are interested in parameter estimation, the Bayesian evidence is merely a normalisation constant and will be dropped in the following. We take the priors specified in Table \ref{tabPars} and implement a binned likelihood defined as
\begin{equation}
\mathcal{L}(\theta) = \prod_i{ \prod_b{  \frac{N_{i,b}^{\bar{N}_{i,b}} }{ \bar{N}_{i,b}! } \exp\left(-N_{i,b}\right)   }} \quad ,
\end{equation}
in which $i$ indexes the experimental setup associated to each target material, $b$ indexes the (ten) energy bins in each experimental setup, $N_{i,b}$ is the number of counts in bin $b$ and target $i$ for the parameter set $\theta$ (computed using equation \eqref{NE1E2}) and $\bar{N}_{i,b}$ is the corresponding number in the mock data (also computed with equation \eqref{NE1E2} but taking the true parameter set). Several comments are in order here. Firstly, we assume negligible background events, which is optimistic but appears reasonable in light of the prospects for low-background, ton-scale direct detection experiments \cite{darwin_IDM2010,darwin,eureca}. It would be interesting to study the impact of background in the findings presented here and, in particular, to understand how target complementarity can help interpreting a WIMP signal in the presence of background. A detailed study including background modelling is left for future work. Secondly, we use the binned likelihood because it is perhaps the most straightforward way to derive constraints without the need to worry about realisation noise. In other words, our ``true'' counts are directly drawn from the corresponding event rate without including Poisson scatter. Therefore our results do not include possible realisation noise; this effect has been included in other works in the literature \cite{Akrami1,Akrami2,APeter2011} where the unbinned likelihood was implemented.

\begin{table*}[htp]
\centering
\fontsize{11}{11}\selectfont
\begin{tabular}{l||llll}
\hline
\hline
Parameter & range & prior & $\quad$ &fiducial \\
\hline
$\textrm{log}_{10} \left( m_{\chi}/\textrm{GeV} \right)$ 		& $(0.1,3.0)$ 	& flat & & $\log_{10} 50$, $\log_{10} 25$, $\log_{10} 250$ \\
$\textrm{log}_{10} \left( \sigma_{p}^{SD}/\textrm{pb}\right)$ 	        & $(-7,-3)$, $(-5,-1)$ 	& flat & & $-5$, $-3$   \\
$a_n/a_p$                                                               & $(-2,2)$      & flat & & $-1$, $0$, $+1$ \\
$\textrm{log}_{10} \left( \sigma_{p}^{sc}/\textrm{pb}\right)$ 	        & $(-11,-7)$, $(-7,-3)$ 	& flat & & $-9$, $-5$   \\
$f_n/f_p$                                                               & $(-4,4)$      & flat & & $1.0$ \\
$\textrm{log}_{10} \left( \sigma_{n}^{vec}/\textrm{pb}\right)$ 	        & $(-11,-7)$ 	& flat & & $-9$   \\
$b_p/b_n$                                                               & $(-2,2)$      & flat & & $0.0$ \\
$\delta\,[\textrm{keV}]$                                                & $(0,200)$     & flat & & $0$, $40$, $100$ \\
\hline
$\rho_0\,[\textrm{GeV/cm}^3]$		& $(0.001,0.9)$	& gaussian: $0.4 \pm 0.1$ & & 0.4 \\
$v_0\,[\textrm{km/s}]$			& $(80,380)$    & gaussian: $230 \pm 30$  & & 230 \\
$v_{esc}\,[\textrm{km/s}]$		& $(379,709)$   & gaussian: $544 \pm 33$  & & 544 \\
$k$		  			&  $(0.5,3.5)$	& flat                    & & 1.0 \\

\hline
\end{tabular}
\caption{\fontsize{9}{9}\selectfont The phenomenological WIMP-related parameters considered to explore the capabilities of the next generation of direct detection instruments. The second and third columns specify the range and prior fed into the MultiNest code, while the last column indicates the fiducial values.}\label{tabPars}
\end{table*}

\par Regarding the astrophysical model -- encoded in the parameters $\rho_0$, $v_0$, $v_{esc}$ and $k$ --, three setups will be used: \emph{(i)} \emph{``fixed'' astrophysics}, where we set the astrophysical parameters to their fiducial values shown in Table \ref{tabPars}; \emph{(ii)} \emph{``varying'' astrophysics}, where uncertainties on the local DM density and velocity distribution are taken into account following Ref.~\cite{paperastrodirect}, namely a flat prior on the shape parameter $k$ and $1\sigma$ priors on $\rho_0$, $v_0$ and $v_{esc}$ according to Table \ref{tabPars}; and \emph{(iii)} \emph{``flat'' astrophysics}, where $\rho_0$, $v_0$, $v_{esc}$ and $k$ are all varied in their ranges with flat priors. The first approach has been extensively used in the literature and assumes an exact knowledge of the astrophysical setup, which is certainly not the case as of today. The second setup, instead, is a reasonable assessment of present uncertainties on the different astrophysical unknowns (see \cite{paperastrodirect} for a detailed discussion), while the third approach assumes rather poor astrophysical knowledge and is useful to study the constraints that can be placed on the parameters $\rho_0$, $v_0$, $v_{esc}$ and $k$ with direct detection alone.

\par Let us finally point out that, since our phenomenological parameters include cross-sections ($\sigma_p^{SD}$, $\sigma_p^{sc}$, $\sigma_n^{vec}$) and WIMP-nucleon couplings ($a_n/a_p$, $f_n/f_p$, $b_p/b_n$) -- instead of WIMP-quark couplings --, we do not fold in nuclear uncertainties pertaining the spin content of nucleons and the $\pi$-nucleon sigma term. For alternative analyses done in the framework of specific WIMP models, see \cite{EllisDD,Akrami1}.

\section{Results}\label{secres}

\subsection{Spin-independent couplings}\label{secSI}

\par The first question we wish to address is whether ton-scale direct detection experiments will be able to determine scalar-proton and scalar-neutron couplings independently, i.e.~measure both $\sigma_p^{sc}$ and $f_n/f_p$ in equation \eqref{sigmaSI2}. Notice that although the widely studied supersymmetric neutralinos feature $f_p\sim f_n$ \cite{EllisDD}, it is rather easy to build WIMP models with uncorrelated proton and neutron couplings $f_p$ and $f_n$ -- for specific implementations see \cite{FengIso,ChangIso,KangIso}. This means that the WIMP-proton and WIMP-neutron scattering amplitudes may interfere non-coherently and the well-known scaling $dR/dE_R\propto A^2$ for spin-independent searches is no longer valid. For the time being vector and axial couplings are neglected. At fixed DM mass, the normalisation of the recoil spectrum is basically set by the quantity
\begin{equation}\label{Isc}
I_{sc} = \sigma_p^{sc}\left(Z+(A-Z)f_n/f_p\right)^2 \quad .
\end{equation}
Therefore, single-target data cannot break the degeneracy between $\sigma_p^{sc}$ and $f_n/f_p$. Instead, by using targets with distinct ratios $(A-Z)/Z$, one can hope to break the degeneracy and constrain scalar-proton \emph{and} scalar-neutron couplings. However, for heavy nuclei -- the most sensitive to SI scattering -- nuclear stability entails an almost universal ratio $(A-Z)/Z\simeq 1.4$. That is why the authors of Ref.~\cite{EllisDD} suggested the use of target elements with $Z\lesssim 17$ (for which $(A-Z)/Z\sim 1$) along with heavy materials in order to measure $f_n/f_p$.

\begin{figure}%[htp]
 \centering
 \includegraphics[width=8.cm,height=7.cm]{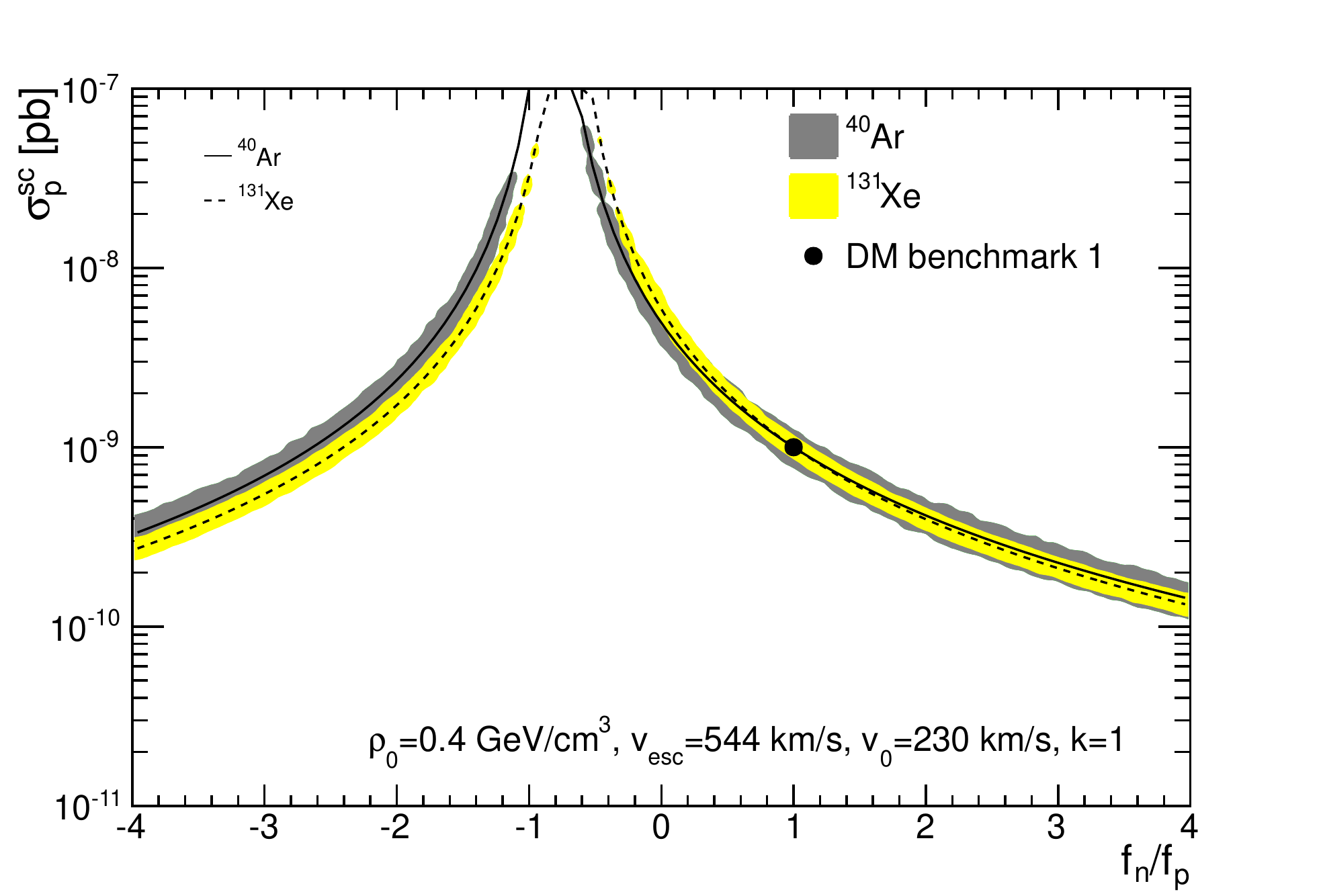}
  \includegraphics[width=8.cm,height=7.cm]{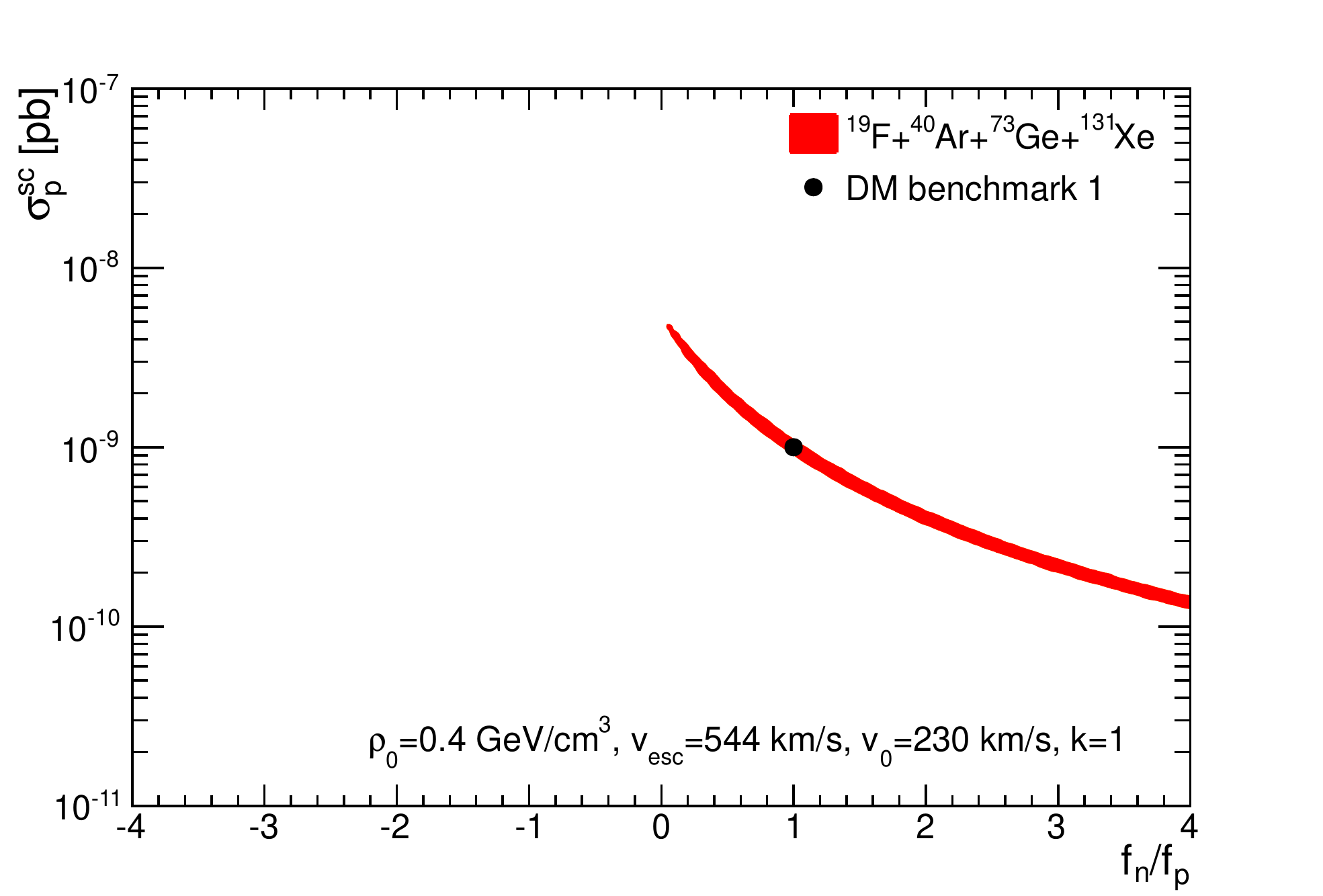}
 \caption{\fontsize{9}{9}\selectfont The joint 95\% posterior probability contours for the DM benchmark 1 and fixing the astrophysical parameters. For simplicity the joint 68\% posterior probability contours are not shown. In the left frame, the grey and yellow contours show the reconstruction capabilities of $^{40}$Ar only and $^{131}$Xe only respectively, while in the right frame the red contour corresponds to $^{19}$F+$^{40}$Ar+$^{73}$Ge+$^{131}$Xe. The solid (dashed) curve in the left frame represents the degeneracy direction according to equation \eqref{Isc} for $^{40}$Ar ($^{131}$Xe). As the Figure shows, a multi-target signal can determine the sign of $f_n/f_p$, but not its absolute value.}\label{figSIpn1}
\end{figure}

\par
\begin{figure}%[htp]
 \centering
 \includegraphics[width=8.cm,height=7.cm]{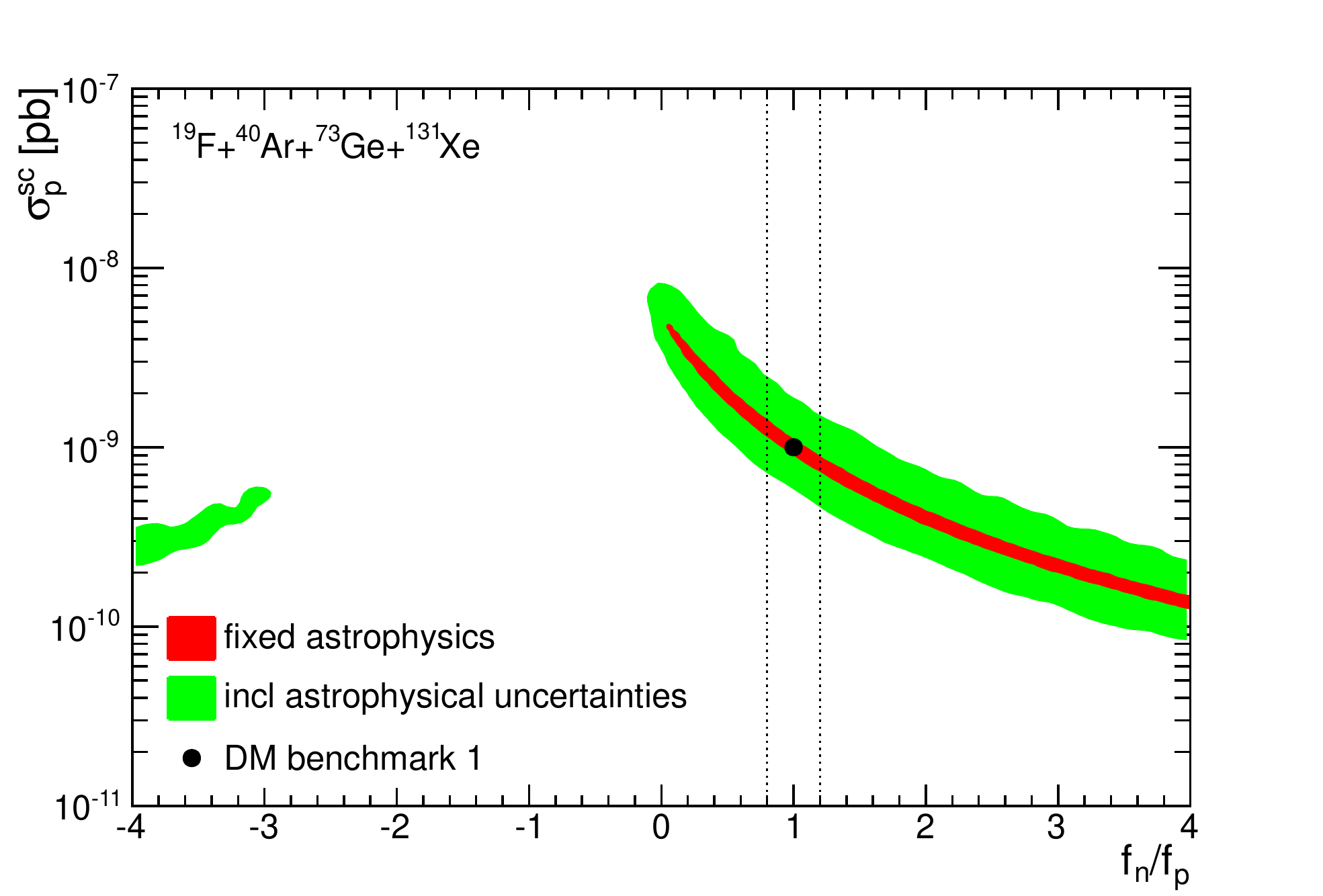} 
 \includegraphics[width=8.cm,height=7.cm]{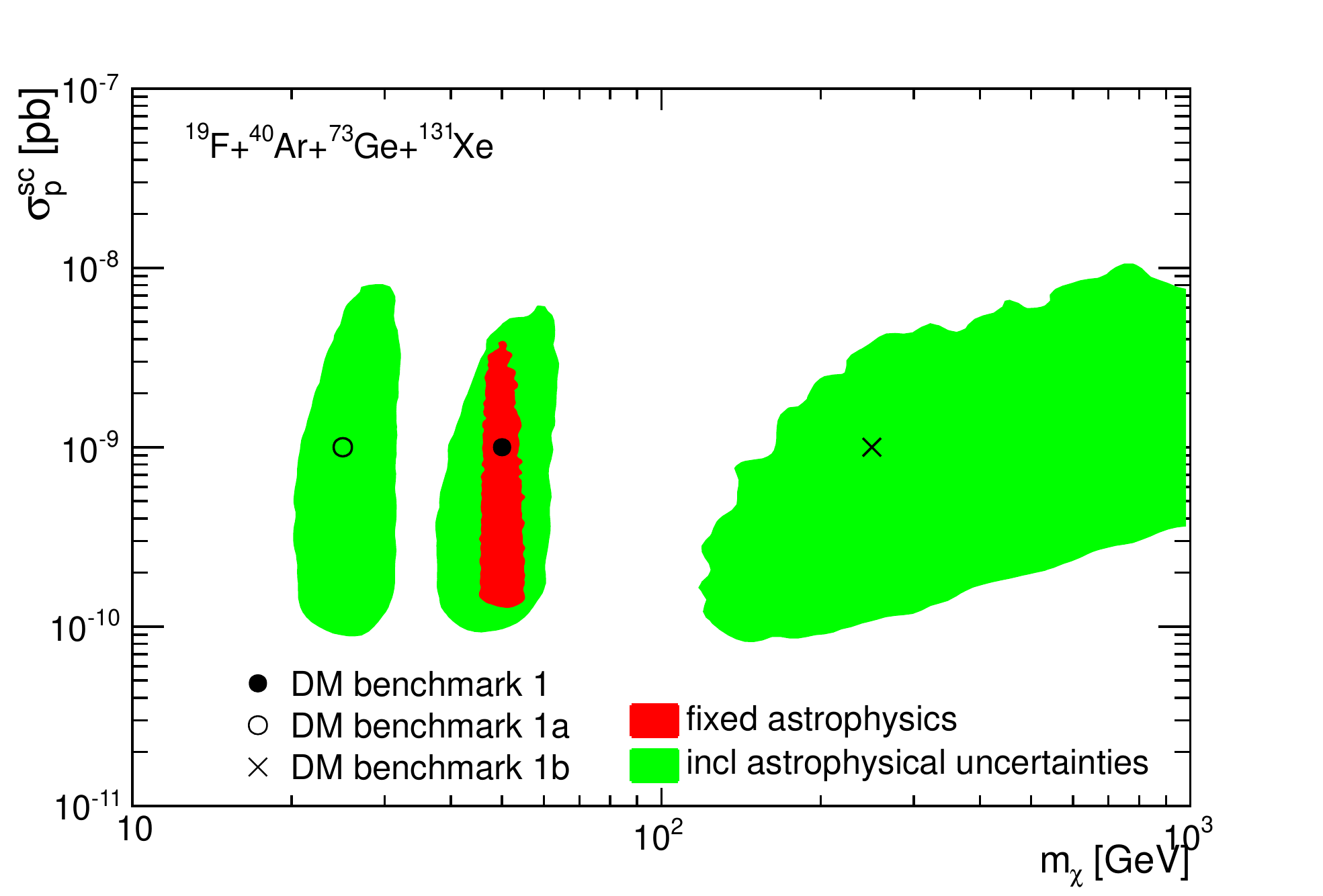}
 \caption{\fontsize{9}{9}\selectfont The joint 95\% posterior probability contours for the DM benchmark 1 and the data set $^{19}$F+$^{40}$Ar+$^{73}$Ge+$^{131}$Xe. The red contours are the same as in Figure \ref{figSIpn1} (right), while the green contours show the effect of marginalising over the astrophysical uncertainties. The vertical dashed lines in the left frame indicate the range $0.8\lesssim f_n/f_p \lesssim 1.2$. Without \emph{a priori} assumptions on the ratio $f_n/f_p$, the next generation of direct detection instruments can only pinpoint $\sigma_p^{sc}$ within 1.5 orders of magnitude (if $m_\chi\sim 50$ GeV). The right plot also shows the 95\% posterior probability contours for the DM benchmarks 1a and 1b when marginalising over astrophysical uncertainties.}\label{figSIpn2}
\end{figure}

 Here, we test such suggestion by taking the DM benchmark 1 in Table \ref{tabBench} as the true model and focussing on $^{19}$F, $^{40}$Ar, $^{73}$Ge and $^{131}$Xe as target materials. Notice that the targets are chosen across the periodic table to maximise their complementarity regarding the ratio $(A-Z)/Z$. Following the procedure outlined in Section \ref{secmeth}, we use the mock data to set constraints on the parameter space defined by $m_\chi$, $\sigma_p^{sc}$ and $f_n/f_p$ (all other parameters are kept fixed to their input values corresponding to benchmark 1 in Table \ref{tabBench}). Figure \ref{figSIpn1} shows the joint 95\% posterior probability contours derived with $^{40}$Ar (left frame, grey), $^{131}$Xe (left frame, yellow) and $^{19}$F+$^{40}$Ar+$^{73}$Ge+$^{131}$Xe (right frame, red) keeping the astrophysical parameters fixed. The solid (dashed) line in the left frame is an iso-contour of the quantity \eqref{Isc} in the case of a $^{40}$Ar ($^{131}$Xe) target. It is evident from Figure \ref{figSIpn1} that target complementarity is particularly effective for determining the correct sign of the $f_n/f_p$ ratio. Nevertheless, different targets constrain essentially the same parameter space for positive values of $f_n/f_p$, and the degeneracy between $\sigma_p^{sc}$ and $f_n/f_p$ cannot be entirely lifted. This degeneracy leads to a 2$\sigma$ uncertainty on $\sigma_p^{sc}$ of about 1.5 orders of magnitude (i.e.~$8\times10^{-11}\textrm{ pb}\lesssim \sigma_p^{sc}\lesssim 6\times10^{-9}\textrm{ pb}$), as can be better appreciated from Figure \ref{figSIpn2} where the effect of marginalising over the astrophysical parameters is shown by the green contours. Notice that the flat bottom-end of the contours in the right plot of Figure \ref{figSIpn2} would extend to lower $\sigma_p^{sc}$ if the prior on $f_n/f_p$ was not restricted to values smaller than 4. The prior on the ratio $f_n/f_p$ (check Table \ref{tabPars}) was intentionally stretched beyond the values found in the literature in order to assess what direct detection experiments can really tell us about this parameter independently of theoretically motivated assumptions. An even more conservative approach would be to adopt a log prior on $f_n/f_p$ (that assumes no knowledge about the scale), but we feel that the flat prior on this parameter is enough for the purposes of the present work. Let us stress at this point that the poor accuracy on the determination of $\sigma_p^{sc}$ stems precisely from the relaxation of the hypothesis $f_n/f_p=1$ (and not from astrophysical uncertainties). Of course, if one is interested in DM candidates with $f_n\simeq f_p$ such as supersymmetric neutralinos, then the contours are restricted to $0.8\lesssim f_n/f_p \lesssim 1.2$ (including nuclear uncertainties, see \cite{EllisDD}), as shown by the vertical dashed lines in Figure \ref{figSIpn2} (left), and the accuracy on $\sigma_p^{sc}$ is drastically improved to about 0.5 orders of magnitude (i.e.~$5\times10^{-10}\textrm{ pb}\lesssim \sigma_p^{sc}\lesssim 2\times10^{-9}\textrm{ pb}$) when marginalising over the astrophysical parameters. We have checked that the contour tail at negative values of $f_n/f_p$ reduces drastically (but does not entirely disappear) when using the target combination $^{19}$F+$^{23}$Na+$^{40}$Ar+$^{73}$Ge+$^{127}$I+$^{131}$Xe (data III) and marginalising over the astrophysical unknowns.

\par On the other hand, an interesting result of our analysis is the fact that the DM mass can be tightly constrained (if $m_{\chi}\sim 50$ GeV) despite of the $\sigma_p^{sc}-f_n/f_p$ degeneracy and the accuracy attained is in line with the ones found in works assuming $f_n=f_p$ (see e.g.~\cite{paperastrodirect,APeter2011}). Notice that this statement does depend on the true WIMP mass: for comparison, we show on the right plot of Figure \ref{figSIpn2} the constraints for the case of $m_\chi=25$ GeV (benchmark 1a) and $m_\chi=250$ GeV (benchmark 1b). The mass accuracy for the three fiducial masses is similar (slightly better, in fact) to the findings of Ref.~\cite{paperastrodirect}, even though we use here one extra target ($^{19}$F) and let $f_n/f_p$ vary. In particular, the 95\% lower limit on the DM mass for the case $m_\chi=250$ GeV shown in Figure \ref{figSIpn2} (right) is slightly more stringent than in Ref.~\cite{paperastrodirect} given the use of $^{19}$F here that was not considered in that work. The accuracy on $\sigma_p^{sc}$, instead, is much worse than in Ref.~\cite{paperastrodirect} and that is due to the $\sigma_p^{sc}-f_n/f_p$ degeneracy explained above. Finally, let us notice that astrophysical uncertainties -- whose impact is stressed in Figure \ref{figSIpn2} -- hinder the determination of all parameters, but leave qualitatively unchanged all the points discussed above. 

\par Overall, the punchline of this exercise is two-fold. Firstly, the next generation of ton-scale direct detection experiments will be able to determine the sign -- but not the scale -- of $f_n/f_p$. This means it will be virtually impossible to tell apart DM candidates with pure scalar-proton and pure scalar-neutron couplings. However, the upcoming instruments have the capability to test isospin-violating dark matter \cite{ChangIso,KangIso,FengIso,FrandsenIso,DelNobile:2011je} as an explanation of DAMA/LIBRA \cite{DAMA2008,DAMA2010} and CoGeNT \cite{cogent} observations compatible with XENON10/100 \cite{Xenon10,Xenon100,Xenon1002} and CDMS \cite{CDMS2009,cdms10} upper limits -- recall that in Ref.~\cite{FengIso} a value $f_n/f_p\simeq -0.7$ was used and this is falsifiable in light of Figure \ref{figSIpn1}. Let us note that the constraints presented in Figures \ref{figSIpn1} and \ref{figSIpn2} may vary significantly depending on the true value of $f_n/f_p$; a full analysis using several true values for this coupling ratio is beyond the scope of this paper but is an interesting topic for future work. Secondly, it seems evident from Figure \ref{figSIpn2} (right) that direct detection alone is able to extract the DM mass precisely (if $m_{\chi}\sim 50$ GeV) but shall not determine $\sigma_p^{sc}$ to better than one order of magnitude. This is an important guideline for future complementarity studies including both direct detection and accelerator searches.

\vspace{0.5cm}

\par Up to now we have assumed vanishing vector couplings. An interesting question is to know whether scalar-interacting and vector-interacting DM candidates can be distinguished with future direct detection data. Since Majorana particles do not have vector couplings but Dirac particles do, this would be a first step in the determination of the nature of dark matter. Clearly, according to equation \eqref{sigmaSI2}, a DM candidate featuring scalar couplings only ($b_n=b_p=0$) with a given $f_n/f_p$ produces exactly the same recoil spectrum as a particle of the same mass and cross-section with vector couplings only ($f'_n=f'_p=0$) and $b'_n/b'_p=f_n/f_p$. These scenarios cannot be told apart. Instead, we shall focus on candidates with $f_n/f_p=1$ and $b_p/b_n=0$. The former is a very typical value for supersymmetric neutralinos (see \cite{EllisDD}), while the latter is featured by sneutrinos and heavy neutrinos \cite{Jungman}. In this case, target complementarity may play a role in extracting both scalar \emph{and} vector couplings since the rate normalisation (for a given mass $m_\chi$) scales as
\begin{equation}\label{Iscvec}
I_{sc+vec} = \sigma_p^{sc}A^2 + \sigma_n^{vec} (A-Z)^2 \qquad (f_n/f_p=1, b_p/b_n=0) \quad  .
\end{equation}

\begin{figure}%[htp]
 \centering
 \includegraphics[width=8.cm,height=7.cm]{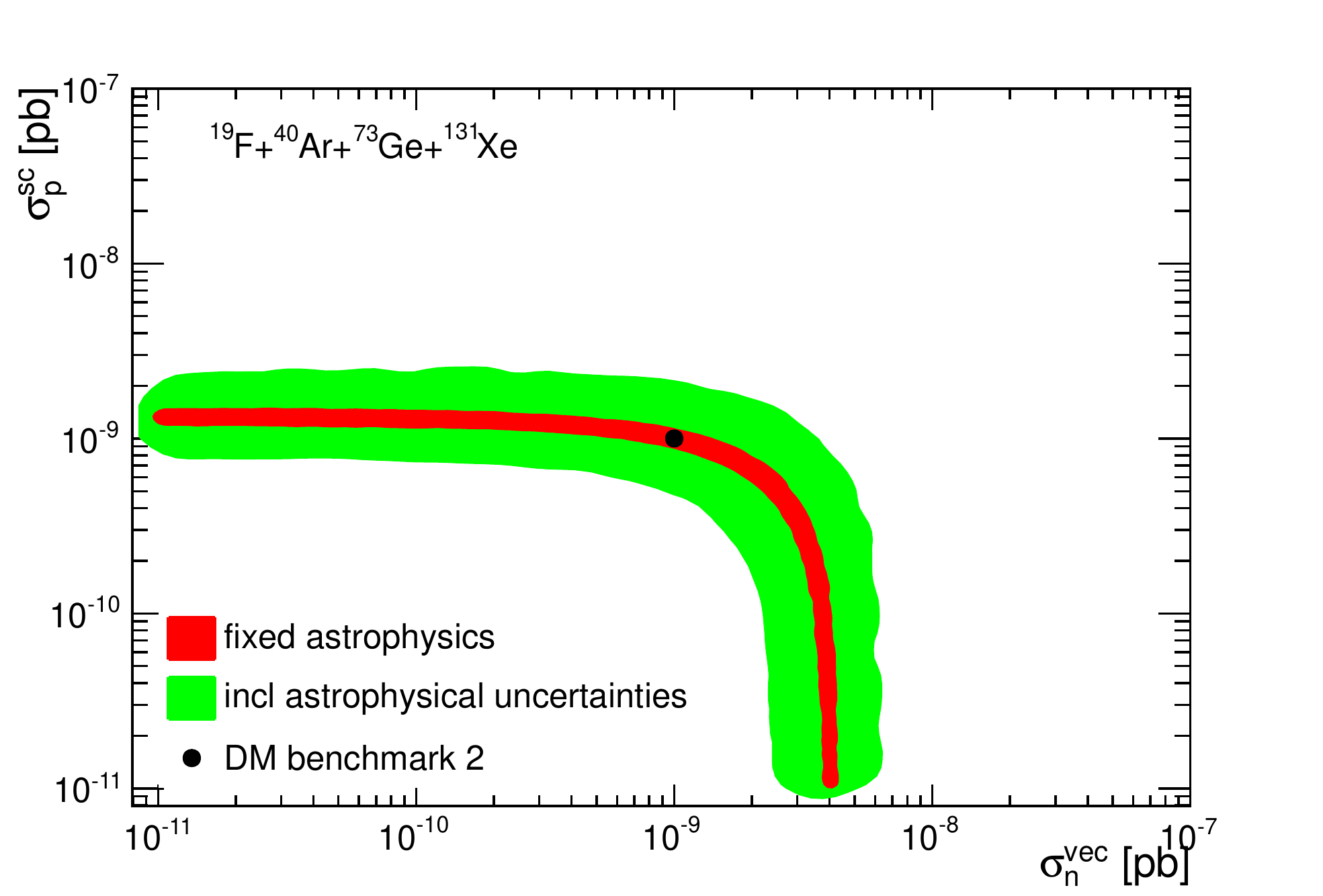} 
 \includegraphics[width=8.cm,height=7.cm]{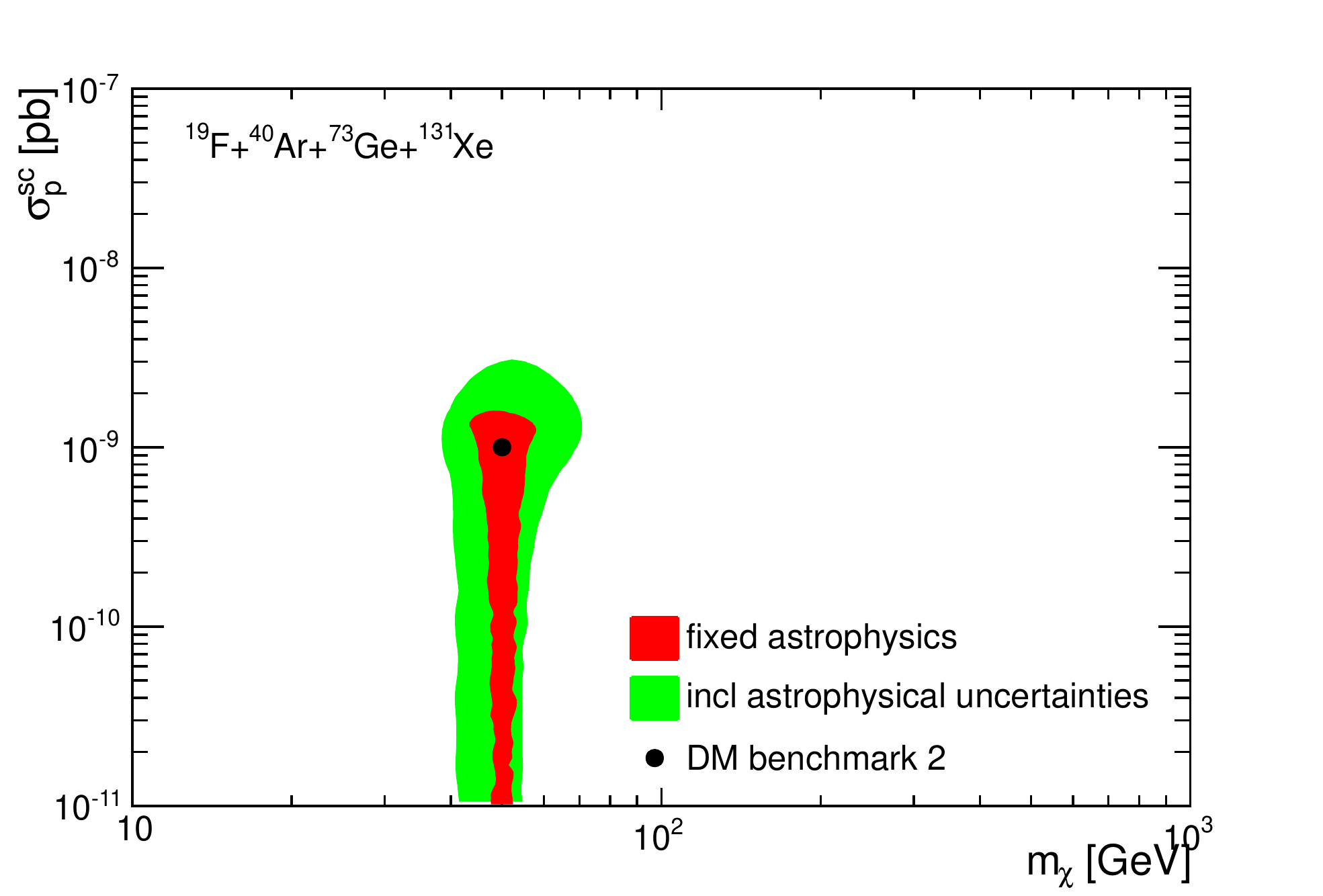} 
 \caption{\fontsize{9}{9}\selectfont The joint 95\% posterior probability contours for the DM benchmark 2 and the data set $^{19}$F+$^{40}$Ar+$^{73}$Ge+$^{131}$Xe. The red contours were obtained for fixed astrophysics, while the green contours include astrophysical uncertainties. Even disregarding astrophysical uncertainties, the distinction between scalar and vector scatterings is virtually impossible.}\label{figSIvec1}
\end{figure}

\par Now, adopting the DM benchmark 2 in Table \ref{tabBench} and following the same procedure as before, we scan over the parameter space $(m_\chi,\sigma_p^{sc},\sigma_n^{vec})$ while fixing $f_n/f_p=1$, $b_p/b_n=0$ and all other parameters to their true values. The joint 95\% posterior probability contours are shown in Figure \ref{figSIvec1} for $^{19}$F+$^{40}$Ar+$^{73}$Ge+$^{131}$Xe (data I) and the cases of fixed (red) and varying (green) astrophysics. As evident from this plot, the prospects for distinguishing between pure scalar, pure vector or mixed DM particles are rather poor. Notice, in particular, that no constraint can be placed on $\sigma_p^{sc}$ nor $\sigma_n^{vec}$ within their priors -- and this statement holds even without marginalising over astrophysical uncertainties. As in the scalar-only scenario, different true values for $f_n/f_p$ and $b_p/b_n$ may lead to different results but such study is deferred to future work.

\subsection{Spin-independent and spin-dependent couplings}\label{secSISD}

\par
\begin{figure}%[htp]
 \centering
 \includegraphics[width=8cm,height=7cm]{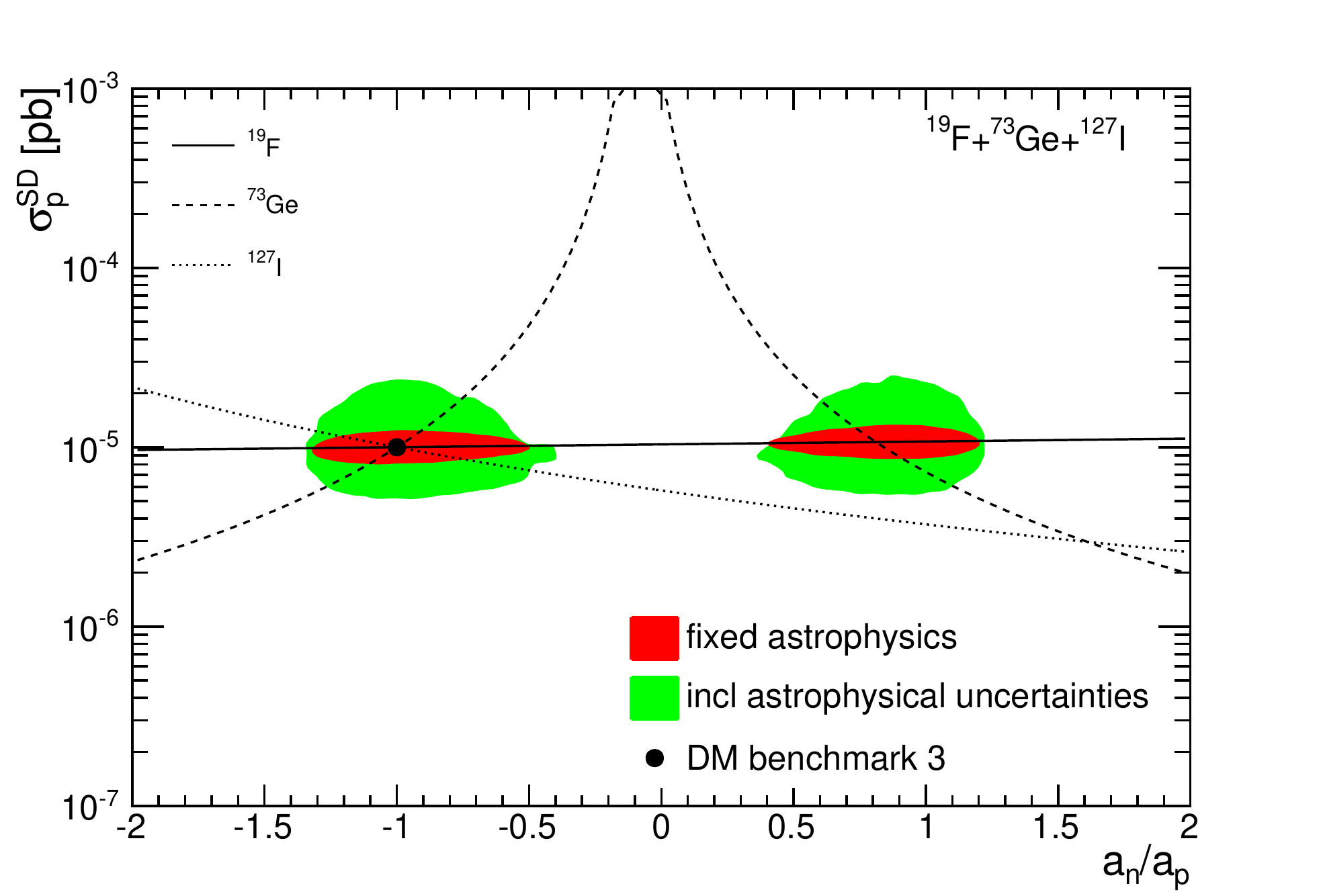} 
 \includegraphics[width=8cm,height=7cm]{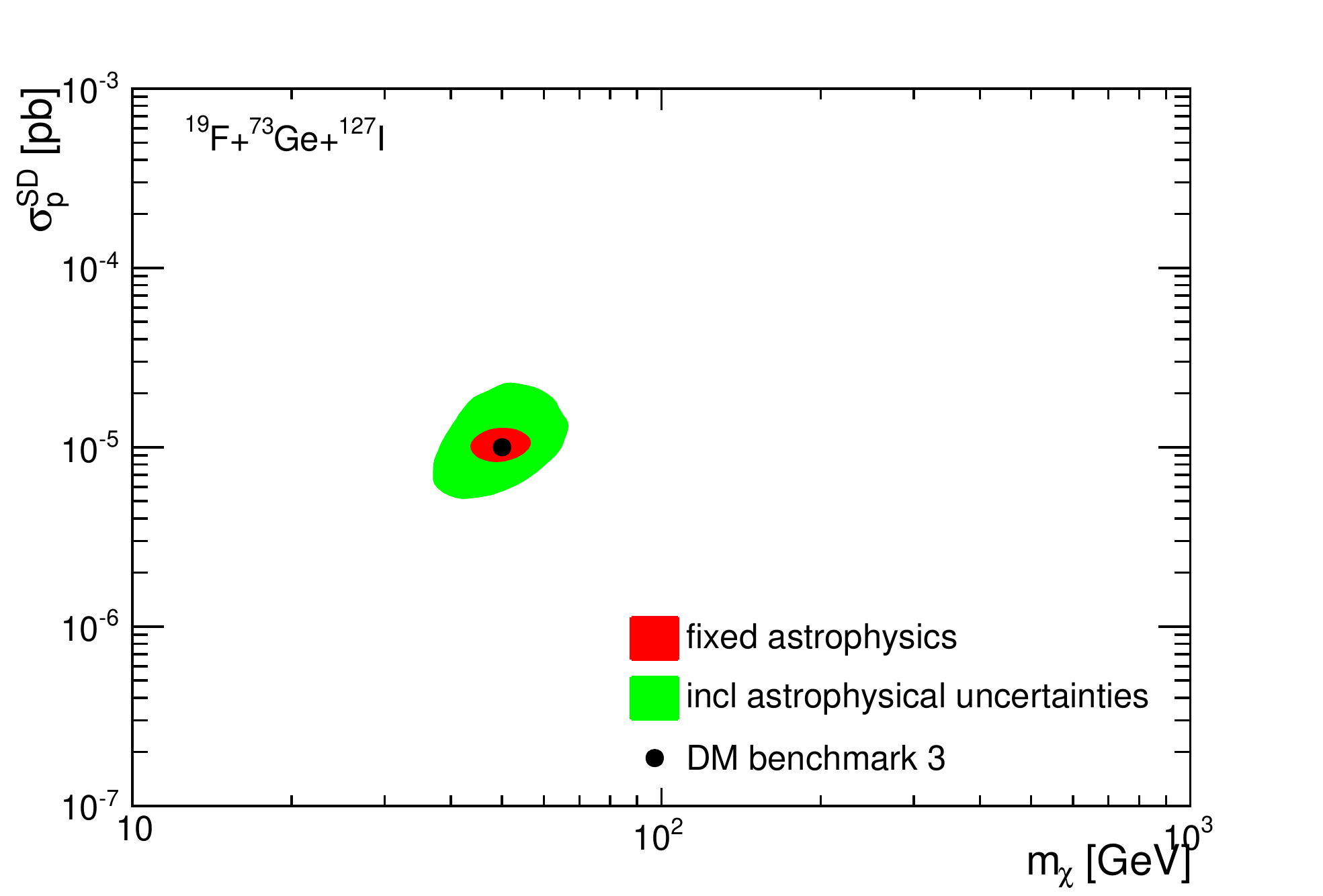} 
 \includegraphics[width=8cm,height=7cm]{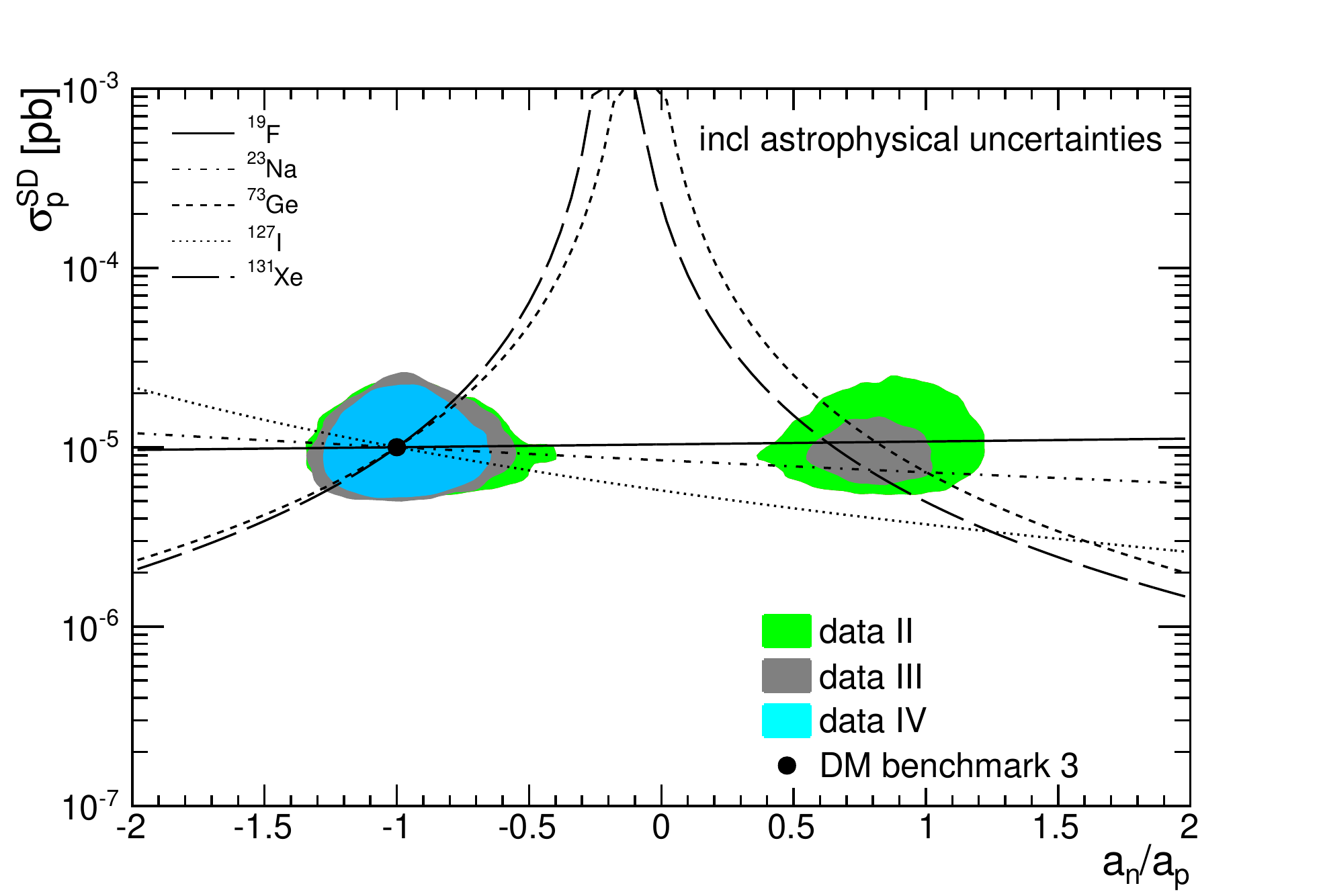}
 \includegraphics[width=8cm,height=7cm]{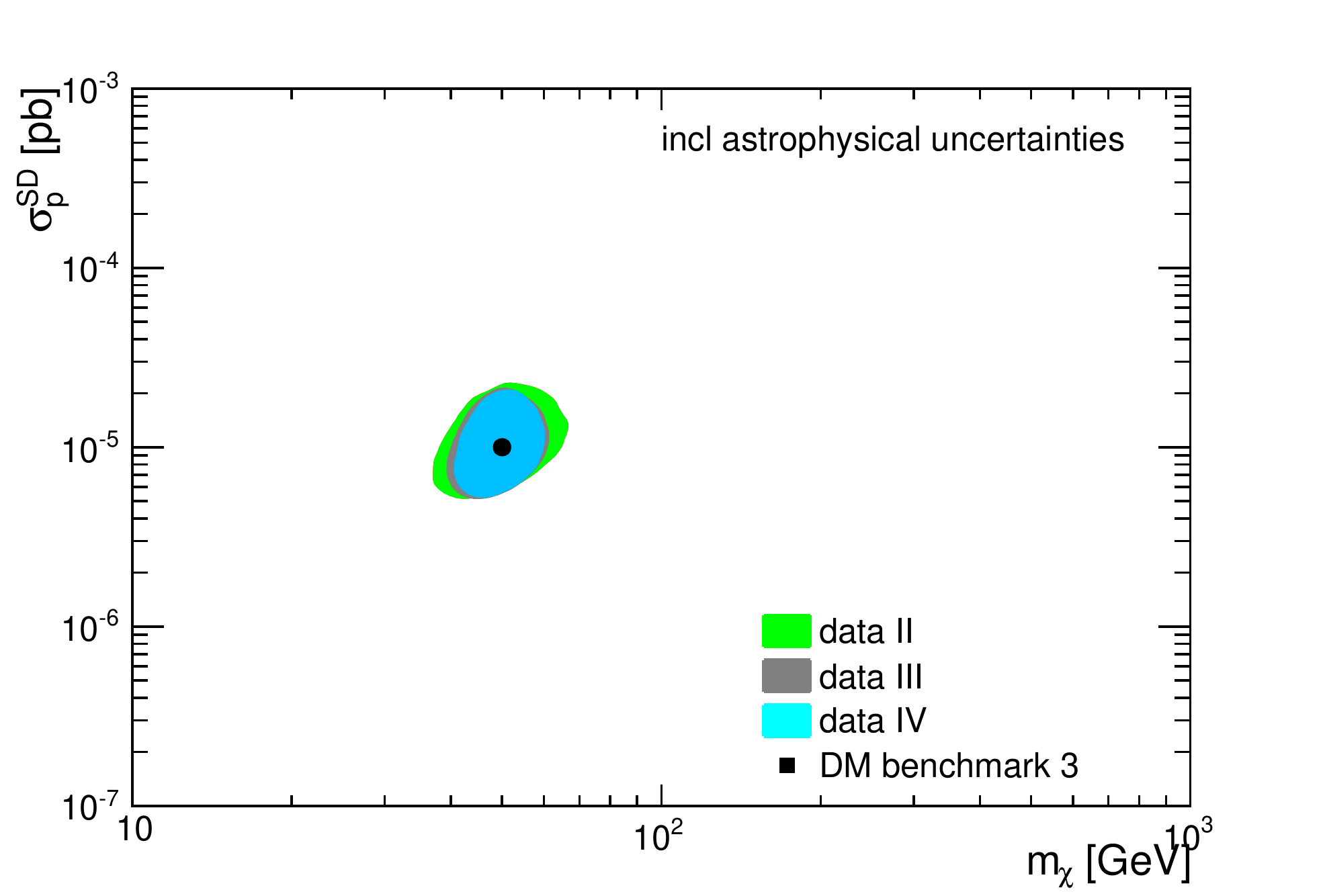}
 \caption{\fontsize{9}{9}\selectfont The joint 95\% posterior probability contours for the DM benchmark 3. In the top frames we show the reconstruction capabilities of $^{19}$F+$^{73}$Ge+$^{127}$I for fixed (red) and varying (green) astrophysics. In the bottom frames we have marginalised over astrophysical uncertainties and show the effect of using different data sets. The curves in the top left and bottom left frames represent the degeneracy directions according to equation \eqref{ISD} for different target nuclei. The ``wrong'' solution at $a_n/a_p>0$ can be discarded by using all target data with an exposure of 3 ton.yr for $^{23}$Na and $^{127}$I.}\label{figSISD1}
\end{figure}

 We now turn to the prospects of measuring SI and SD cross-sections with upcoming direct detection experiments. Despite the $\sigma_p^{sc}-f_n/f_p$ degeneracy discussed in the previous Section, here we shall fix $f_n/f_p=1$ and work in the parameter space $(m_\chi,\sigma_p^{sc},\sigma_p^{SD},a_n/a_p)$. Once again, vector couplings are neglected. This has been the setup used in several works such as \cite{Vergados} or \cite{Akrami1}. As pointed out in Ref.~\cite{Vergados} and clear from equations \eqref{sigmaSD2} and \eqref{sigmaSI2} in Section \ref{secform}, the different couplings can be constrained by using a target with good SI sensitivity and targets with distinct $\langle S_p^N\rangle / \langle S_n^N \rangle$ in order to lift the degeneracy between $\sigma_p^{SD}$ and $a_n/a_p$. Indeed, while the normalisation of the SI rate in a given material (for fixed $m_\chi$) is given by equation \eqref{Isc}, the SD rate depends on 
\begin{equation}\label{ISD}
I_{SD} = \sigma_p^{SD} \left(\langle S_p^N \rangle + \langle S_n^N \rangle a_n/a_p \right)^2 \quad .
\end{equation}
The iso-contours of this quantity for different nuclei are shown in Figure \ref{figSISD1} (top left and bottom left frames) in the case of benchmark 3. The complementarity between the targets is clear and that is the key to measure \emph{simultaneously} $\sigma_p^{SD}$ and $a_n/a_p$ as we shall see. Notice that the DM benchmark 3 features $a_n/a_p=-1$ which is a typical value when considering supersymmetric neutralinos -- in fact, these usually present $a_n>0$ and $a_p<0$ \cite{Vergados,Akrami1}.

\par To begin with, we consider the targets proposed in \cite{Vergados}, i.e.~$^{19}$F+$^{73}$Ge+$^{127}$I. Using the mock counts associated to benchmark 3 and each target material, we shown in the top plots of Figure \ref{figSISD1} the joint 95\% posterior probability contours for $^{19}$F+$^{73}$Ge+$^{127}$I (data II) and the cases of fixed (red) and varying (green) astrophysics. Note that the contours are well inside the $a_n/a_p$ range, which justifies our flat prior on this parameter. Now, the posterior is markedly bimodal: there is a (``correct'') solution for $a_n/a_p<0$ and a (``wrong'') solution for $a_n/a_p>0$. This is precisely what was found in \cite{Vergados} where, using total event rates, the authors identified one solution for each sign of $a_n/a_p$. Our analysis, however, makes use of the full energy spectrum and includes the impact of marginalising over astrophysical uncertainties. Moreover, for fixed astrophysics we anticipate a $2\sigma$ measurement $a_n/a_p=0.85\pm 0.36$ (restricted to $a_n/a_p>0$) and $a_n/a_p=-0.95\pm 0.38$ (restricted to $a_n/a_p<0$) with $^{19}$F+$^{73}$Ge+$^{127}$I. Marginalising over the astrophysical parameters results in very similar $a_n/a_p$ constraints as can be seen by the contours in the top left frame of Figure \ref{figSISD1}. In particular, the absolute value of $a_n/a_p$ can be well constrained, but not its sign. This is essentially opposite to the case of $f_n/f_p$ for which the sign can be determined but not its scale, cf.~Section \ref{secSI} and Figures \ref{figSIpn1} and \ref{figSIpn2}. The reason for such difference lies in the different signs and values of $(A-Z)/Z$ (for $f_n/f_p$) and $\langle S_n^{N} \rangle/\langle S_p^{N}\rangle$ (for $a_n/a_p$) featured by the nuclei at use. Indeed, the values of $(A-Z)/Z$ just slightly vary between 1.11 ($^{19}$F) and 1.43 ($^{131}$Xe), while the ratio $\langle S_n^{N} \rangle/\langle S_p^{N}\rangle$ may be either positive or negative and either very small ($-$0.018 for $^{19}$F) or very large (12.6 for $^{73}$Ge). Also, notice that a true ratio $a_n/a_p\neq -1$ leads in principle to significantly distinct constraints on the WIMP parameter space, as we shall see below. In any case, despite the degeneracy between $\sigma_p^{SD}$ and $a_n/a_p$, we find that $m_\chi$, $\sigma_p^{sc}$ and $\sigma_p^{SD}$ are determined to a good accuracy using simply $^{19}$F, $^{73}$Ge and $^{127}$I for a DM particle with the properties of benchmark 3.

\par For comparison, we focus on Ref.~\cite{Akrami1} where a rather complete study in the context of supersymmetry was presented. The authors have used the projected capabilities of ton-scale xenon, germanium and COUPP-like experiments to constrain the supersymmetric parameter space, including both astrophysical and nuclear uncertainties. For their DM benchmark 1 ($m_\chi\simeq 108$ GeV, $\sigma_p^{sc}\simeq 3.9\times 10^{-9}$ pb, $\sigma_p^{SD} \simeq 2.8 \times 10^{-5}$ pb, $a_n/a_p\simeq -0.87$), they are able to measure all three cross-sections ($\sigma_p^{sc}$, $\sigma_p^{SD}$, $\sigma_n^{SD}$) and find that the 2$\sigma$ uncertainty on $a_p$ and $a_n$ is of order $50\%$ -- check their top right plot in Figure 11. It is reassuring that this uncertainty is in line with the values stated in the last paragraph, although the analyses are not strictly comparable. Let us notice, however, that the results in Ref.~\cite {Akrami1} are basically restricted to $a_p<0, \, a_n>0$ (or, in our notation, $a_n/a_p<0$) because the authors are working on the supersymmetric framework, which means that a possible solution at $a_n/a_p>0$ (our right blob in the top left plot of Figure \ref{figSISD1}) is dismissed \emph{a priori}.

\par
\begin{figure}%[htp]
 \centering
 \includegraphics[width=8cm,height=7cm]{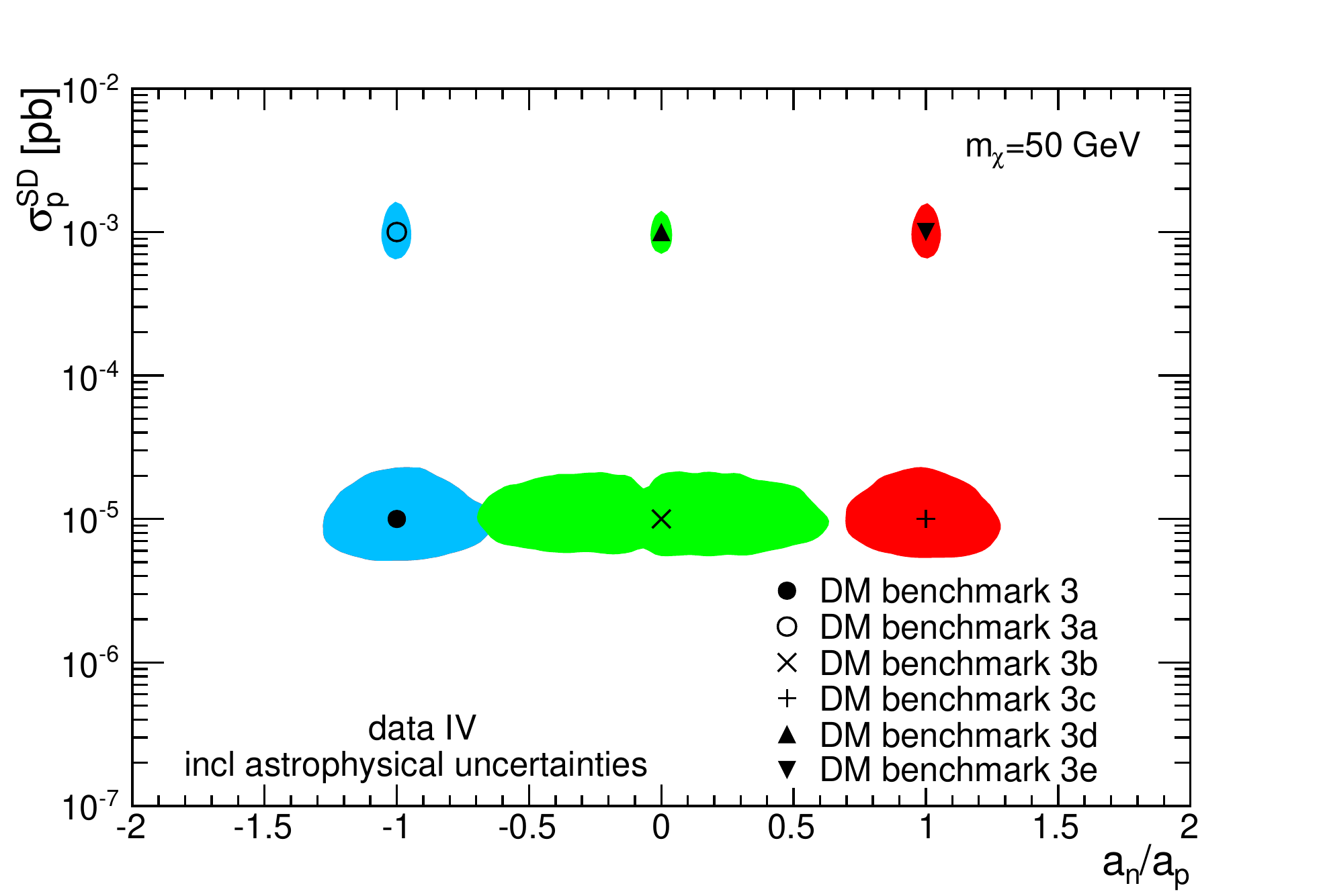} 
 \includegraphics[width=8cm,height=7cm]{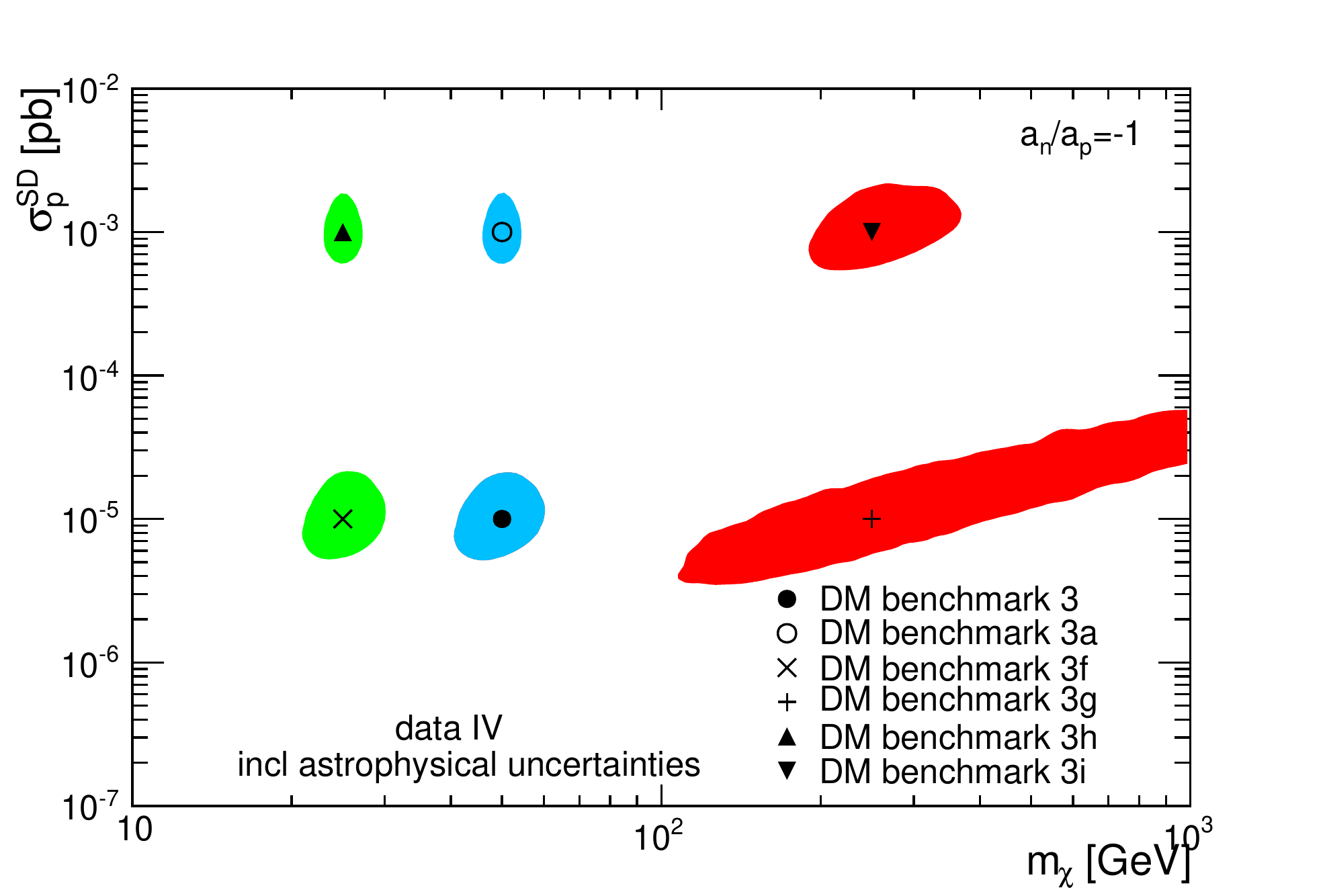}  
 \caption{\fontsize{9}{9}\selectfont The joint 95\% posterior probability contours for the DM benchmarks 3 and 3a--i using data IV and marginalising over the astrophysical parameters. The plots show how the accuracy in the extraction of WIMP parameters can change dramatically according to the true underlying WIMP properties.}\label{figSISD2}
\end{figure}

 In order to test if further direct detection data are able to break the $\sigma_p^{SD}-a_n/a_p$ degeneracy in Figure \ref{figSISD1} (top frames), we have rerun the scan but now making use of all targets, $^{19}$F+$^{23}$Na+$^{40}$Ar+$^{73}$Ge+$^{127}$I+$^{131}$Xe (data III). The corresponding posterior probability contours are displayed in grey in the bottom plots of Figure \ref{figSISD1}, where we have marginalised over astrophysical uncertainties. The wrong solution -- at $a_n/a_p>0$ -- is now more tightly constrained, but cannot be excluded at the 2$\sigma$ level. This persistence of the dual solution is due to the fact that three target nuclei ($^{19}$F, $^{73}$Ge and $^{131}$Xe) are approximately degenerate in the regions around $a_n/a_p\simeq \pm1$ as shown by the line contours in Figure \ref{figSISD1} (bottom left frame). The dual-shaped posterior does depend on the true ratio $a_n/a_p$ (here $a_n/a_p=-1$) and its exact behaviour is in principle different for distinct true $a_n/a_p$ values -- that is an interesting topic that we leave for future research (cf.~also Figure \ref{figSISD2}). Eventually, to completely discard the posterior volume at $a_n/a_p>0$, one can bet on a larger exposure for targets like $^{23}$Na and $^{127}$I that effectively break through the degeneracy. That is what we have done by considering a fourth data set including all targets as in data III but assuming enhanced $^{23}$Na and $^{127}$I exposures, $\epsilon_{eff}(^{23}\textrm{Na})=\epsilon_{eff}(^{127}\textrm{I})=3$ ton.yr. The results are shown in the bottom plot of Figure \ref{figSISD1} by the cyan contours and entail the $2\sigma$ measurements $\log_{10}(\sigma_p^{SD}/\textrm{pb})=-5.0\pm 0.25$ or $\sigma_p^{SD}=1.0^{+0.78}_{-0.44}\times10^{-5}$ pb and $a_n/a_p=-0.97\pm 0.40$ when marginalising over the astrophysical parameters. In this case, the positive $a_n/a_p$ solution is entirely ruled out (at $\sim 10\sigma$). This means that both $\sigma_p^{SD}$ and $\sigma_n^{SD}$ (or equivalently $\sigma_p^{SD}$ and $a_n/a_p$) are measurable with future direct detection data.

\par Up to now we have focussed on the DM benchmark 3 featuring $m_\chi=50$ GeV, $\sigma_p^{SD}=10^{-5}$ pb and $a_n/a_p=-1$. It is however important to check how our results depend on the true DM mass $m_\chi$ and SD coupling ratio $a_n/a_p$. Also, the latest experimental limits \cite{Felizardo:2010mi,Felizardo:2011uw,XENONSD} still allow SD cross-sections as large as $10^{-3}$ pb. Therefore, the additional DM benchmarks 3a--i in Table \ref{tabBench} are studied and the corresponding 95\% posterior probability contours using data IV and marginalising over astrophysical unknowns are shown in Figure \ref{figSISD2}. As can be seen in the left frame of this Figure, the $a_n/a_p$ uncertainty is a complicated function of the underlying true value of $a_n/a_p$ and appears larger for true values $a_n/a_p\sim 0$ than $|a_n/a_p|\sim 1$ (when $\sigma_p^{SD}=10^{-5}$ pb). Moreover, for the DM benchmarks 3a, 3d and 3e with $\sigma_p^{SD}=10^{-3}$ pb, the uncertainties shrink considerably given that the fraction of SD events relative to SI events is much larger than for $\sigma_p^{SD}=10^{-5}$ pb. In the right frame of the same Figure one can appreciate the reconstruction capabilities for different values of $m_\chi$ and $\sigma_p^{SD}$ -- as expected, high masses and low cross-sections lead to larger uncertainties.

\subsection{Inelastic parameter}\label{secSId}

\begin{figure}%[htp]
 \centering
 \includegraphics[width=8.cm,height=7.cm]{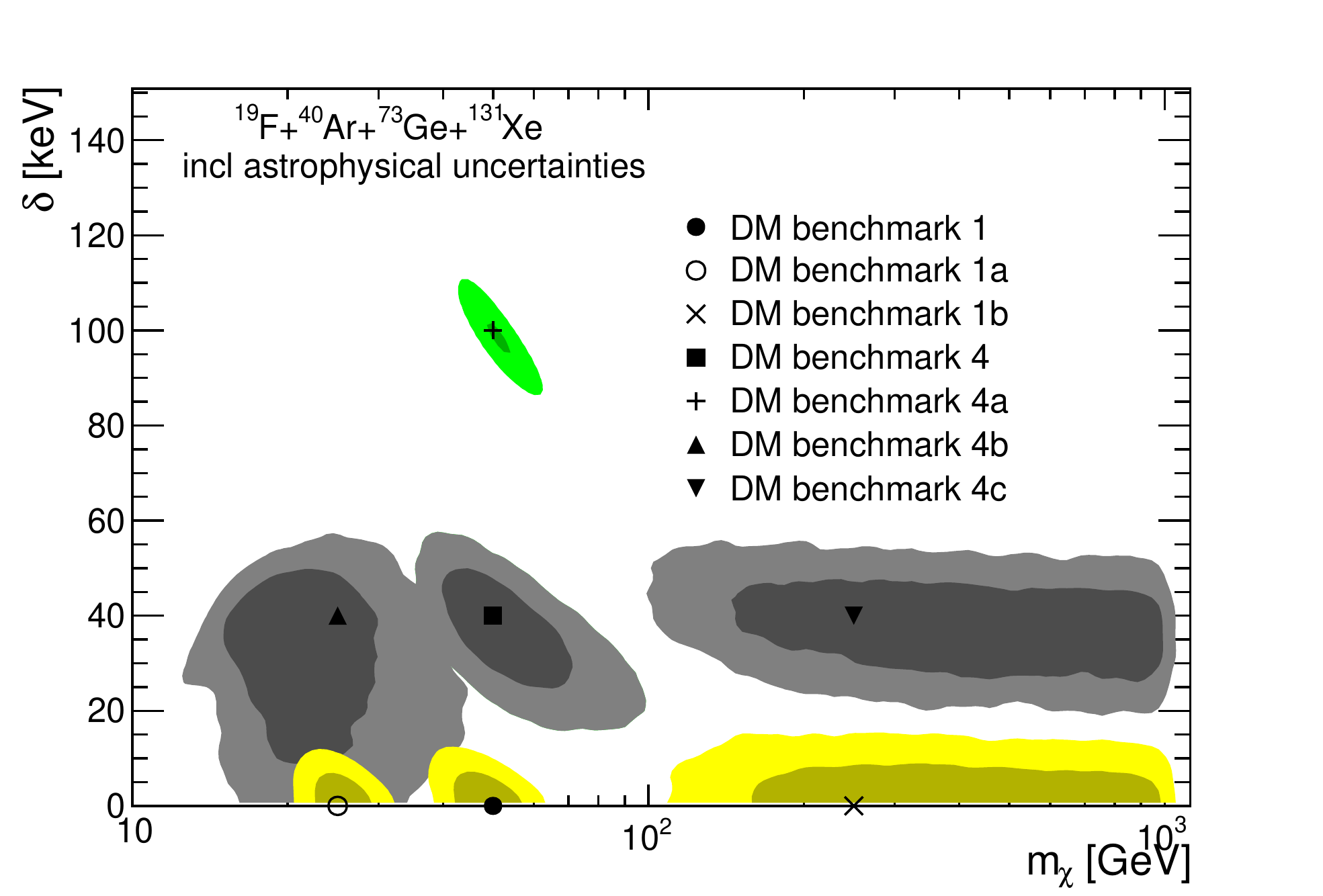} 
 \includegraphics[width=8.cm,height=7.cm]{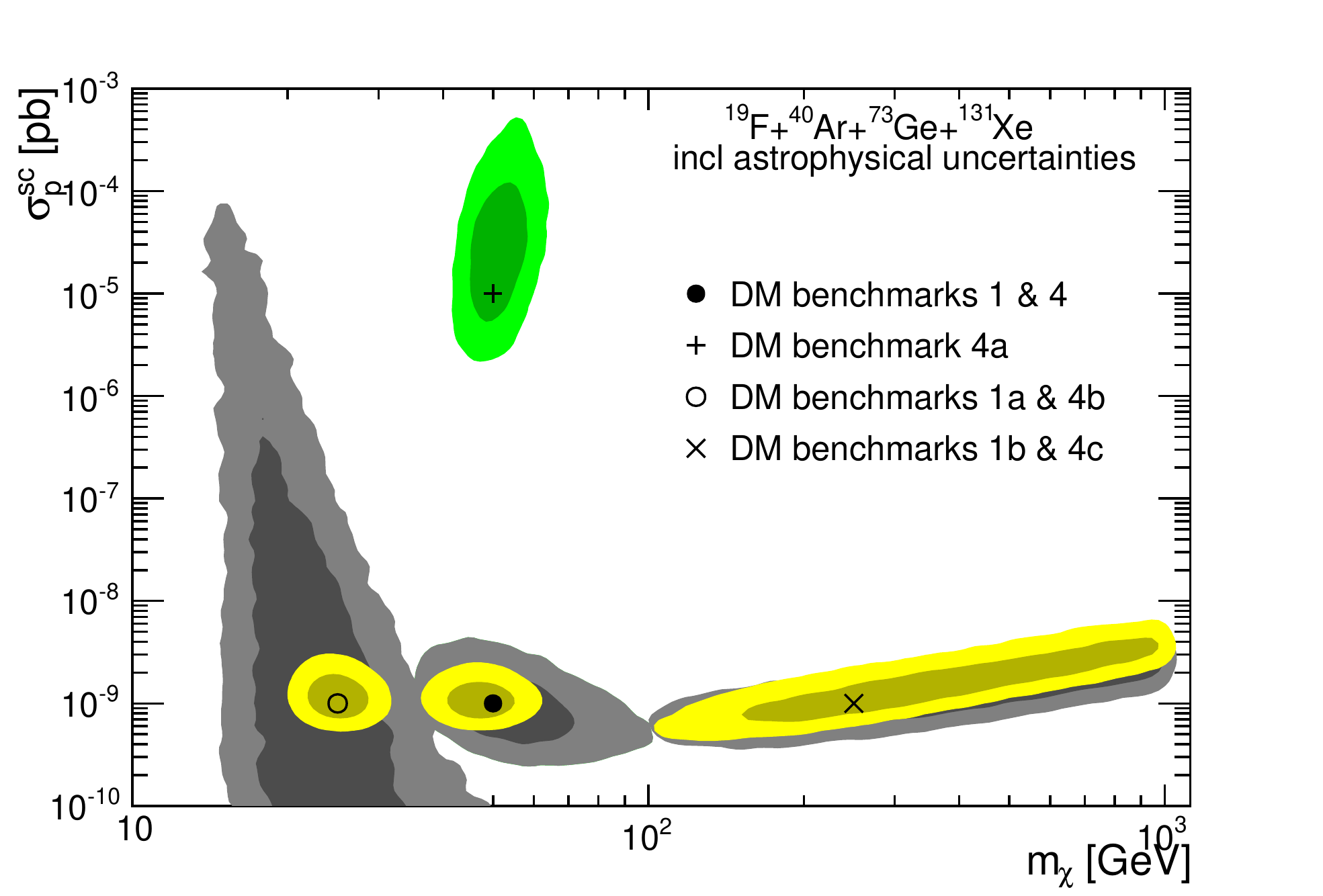} 
 \caption{\fontsize{9}{9}\selectfont The joint 68\% and 95\% posterior probability contours for the data set $^{19}$F+$^{40}$Ar+$^{73}$Ge+$^{131}$Xe and marginalising over the astrophysical uncertainties. The different contours were obtained assuming the DM benchmark 1, 1a, 1b, 4, 4a, 4b and 4c. In the left frame the constraints are adjacent to the corresponding DM benchmark and in the right frame the same colour coding is used. The left plot shows that the next generation of instruments will be able to determine the inelastic parameter $\delta$ with an accuracy of tens of keV.}\label{figSId1}
\end{figure}

\par In direct dark matter searches, the WIMP-nucleus scattering is widely assumed to be elastic when deriving constraints on the WIMP properties. This basically amounts to setting $\delta=0$ in equation \eqref{vmin}. However, as long noticed in the literature \cite{InelasticWeiner}, inelastic dark matter scenarios are perfectly viable and lead to a very diversified phenomenology. In fact, while in the elastic case ($\delta=0$) $v_{min}\propto\sqrt{E_R}$ and thus the recoil rate rapidly decays with increasing energy, inelastic dark matter ($\delta>0$) induces a behaviour $v_{min}\propto 1/\sqrt{E_R}$ ($v_{min}\propto \sqrt{E_R}$) at sufficiently small (large) recoil energies which translates into a peak in the event rate (damped by the nuclear form factor). Inelastic DM models featuring $\delta\simeq 0-100$ keV have been proposed \cite{InelasticWeiner} to reconcile DAMA/LIBRA findings with the null results from other direct detection experiments, although these models are now in tension with present data \cite{Angle:2009xb,Ahmed:2010hw,Xenon100Ine}. It is therefore an important matter to study the prospects of measuring -- or at least constraining -- the inelastic parameter $\delta$ with future ton-scale instruments. Ref.~\cite{fox2} has studied how $\delta$ (and $m_\chi$) can be constrained independently of astrophysics. Here, we pursue a slightly more aggressive approach by modelling astrophysical uncertainties as outlined in Section \ref{secmeth}, and present the expected accuracy on $\delta$ once the next generation of instruments comes online. For definiteness, we assume the DM benchmarks 1, 1a, 1b ($\delta=0$), 4, 4b, 4c ($\delta=40$ keV) and 4a ($\delta=100$ keV), while scanning over $(m_\chi,\sigma_p^{sc},\delta)$ and fixing all other parameters to their true values. Notice that for the DM benchmark 4a with $\delta=100$ keV we use an enhanced scalar cross-section $\sigma_p^{sc}=10^{-5}$ pb (still perfectly compatible with null searches for $\delta=100$ keV and $m_\chi=50$ GeV, cf.~Figure 4 in Ref.~\cite{Angle:2009xb}) in order to have a non-negligible event rate for the used targets. Figure \ref{figSId1} shows the joint 68\% and 95\% posterior probability contours for the different DM benchmarks when using $^{19}$F+$^{40}$Ar+$^{73}$Ge+$^{131}$Xe (data I) and marginalising over the astrophysical parameters. It is remarkable that the complementarity between different targets enables the measurement of $\delta$ rather accurately. In particular, we obtain a $2\sigma$ range $\delta=4 \pm 6$ keV for benchmark 1, $\delta=36 \pm 17$ keV for benchmark 4 and $\delta= 98 \pm 11$ keV for benchmark 4a, all featuring $m_\chi=50$ GeV. Moreover, similar results are obtained for the DM benchmarks with $m_\chi=25$ GeV (1a, 4b) and $m_\chi=250$ GeV (1b, 4c) as evident from Figure \ref{figSId1}. This is a very interesting result indeed: even in the case of benchmarks featuring $m_\chi=250$ GeV, for which the DM mass itself can only be bounded from below, the inelastic parameter $\delta$ can be robustly pinpointed. Recall that the results in this Section are obtained assuming $f_n/f_p=1$, so care must be taken in interpreting these figures in light of the discussion in Section \ref{secSI}. Finally, we have checked that there is a significant correlation between $\delta$ and $v_0$, so that a $v_0$ measurement tighter than presented in Table \ref{tabPars} may lead in the future to a better determination of the inelastic parameter.

\subsection{General case}\label{secall}

\par
\begin{figure}%[htp]
 \centering
 \includegraphics[width=8.cm,height=7.cm]{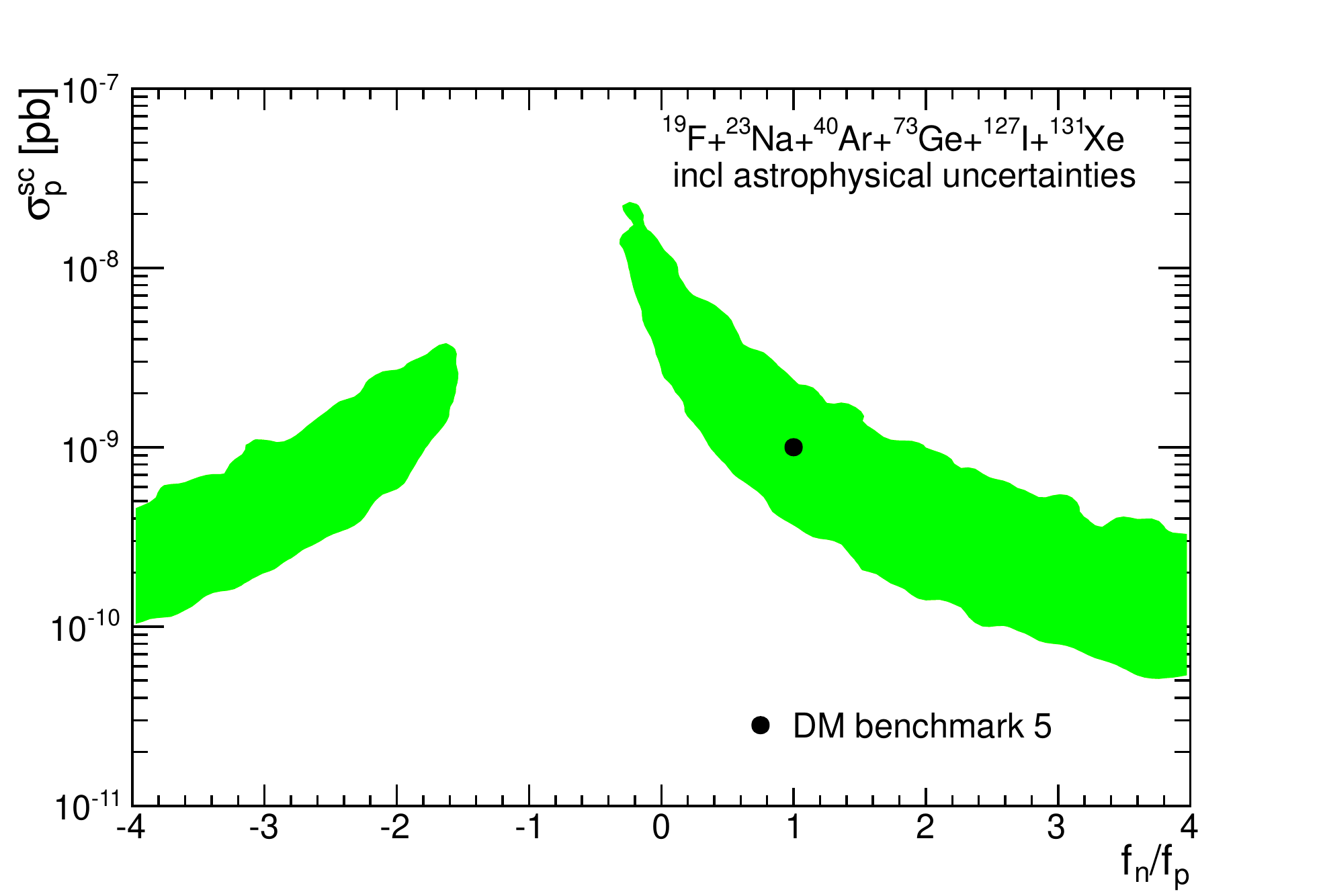} 
 \includegraphics[width=8.cm,height=7.cm]{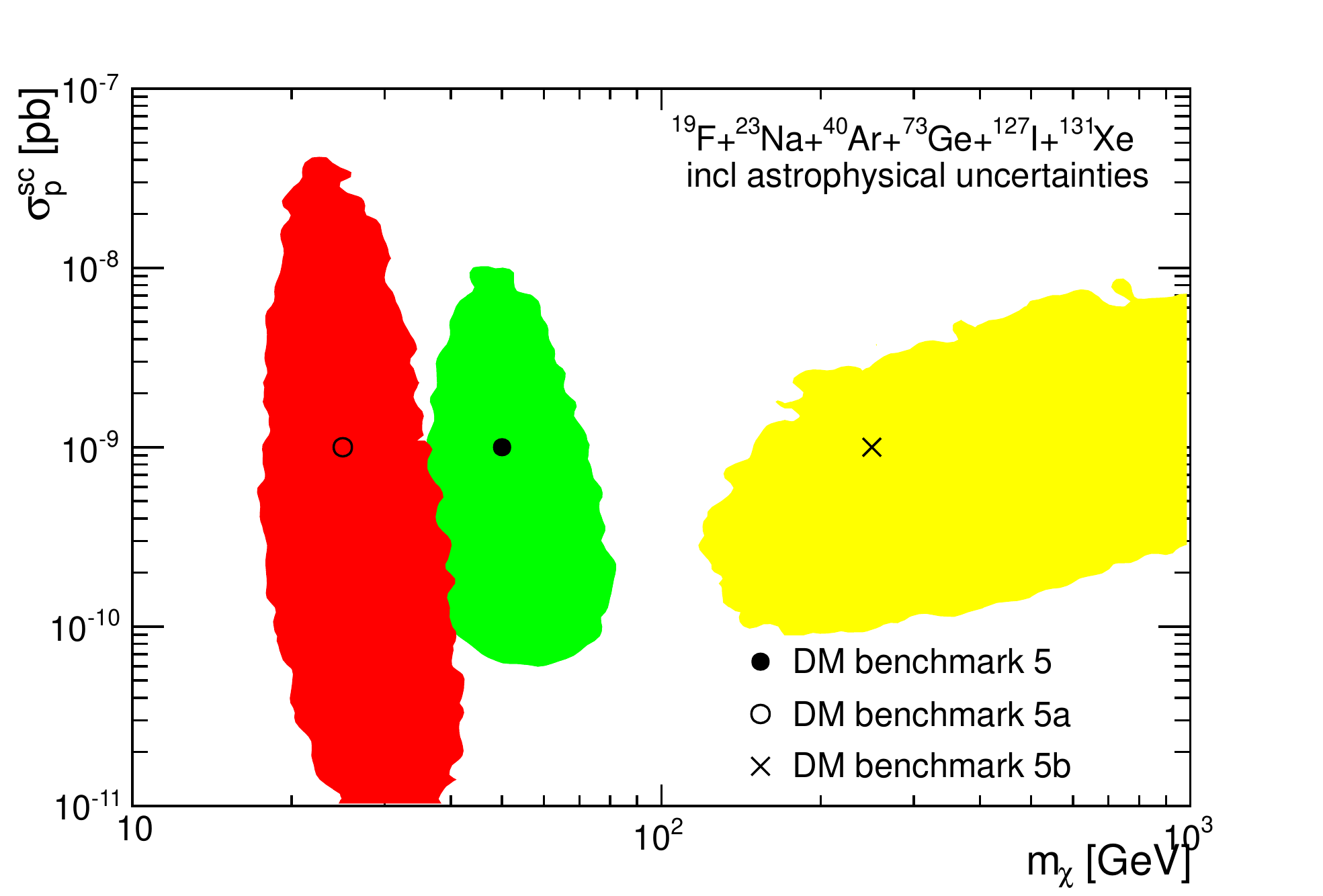} 
  \includegraphics[width=8.cm,height=7.cm]{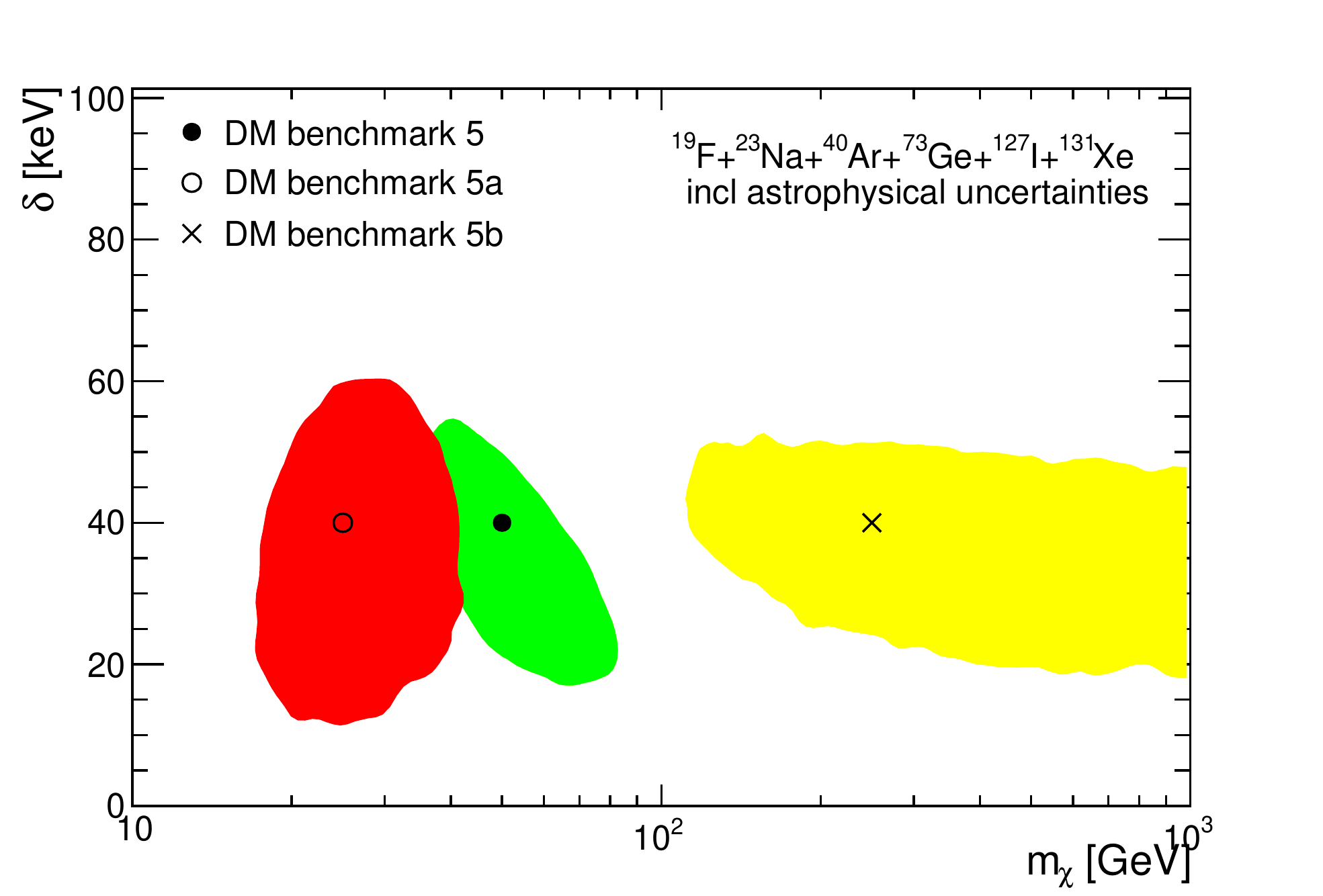} 
 \includegraphics[width=8.cm,height=7.cm]{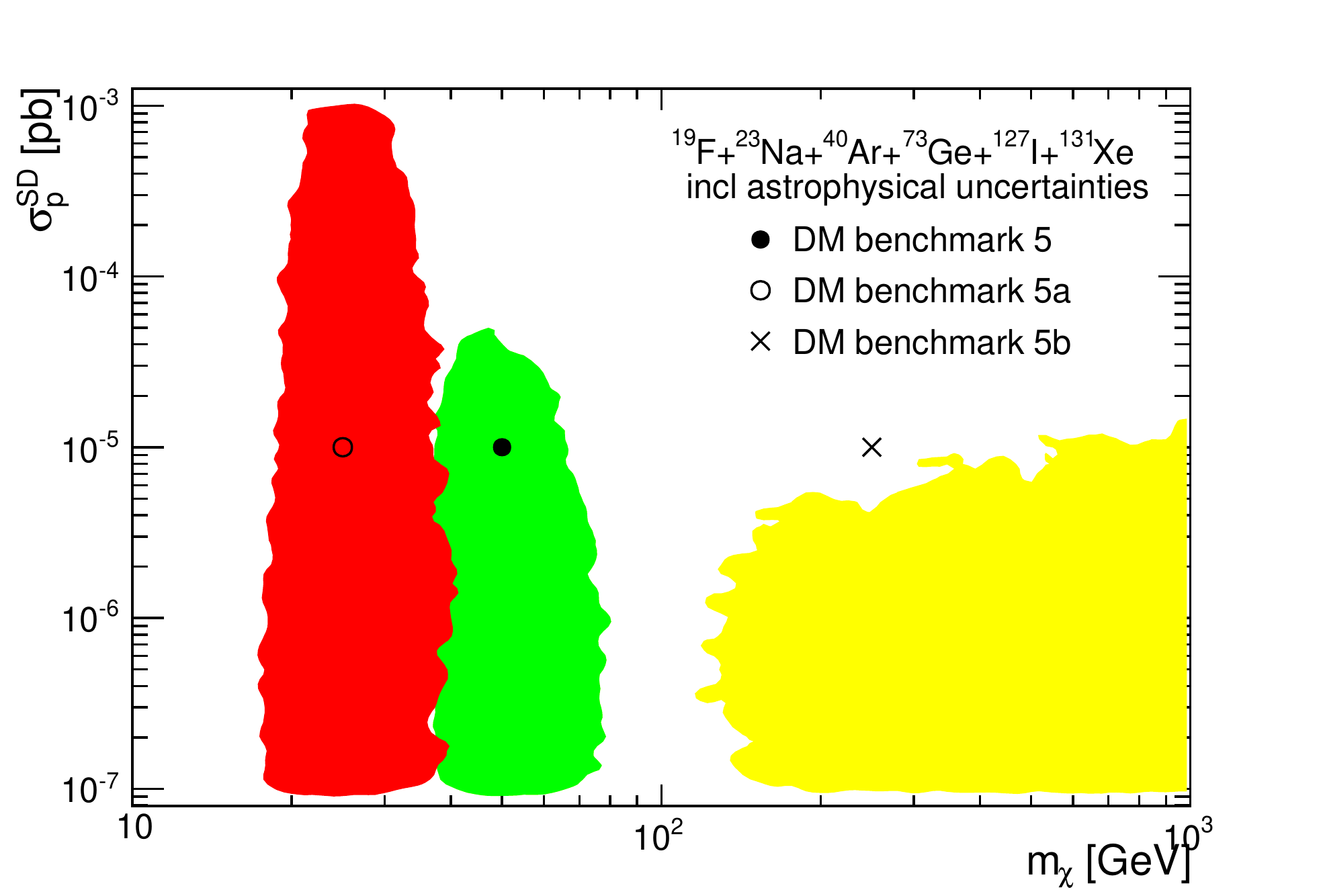} 
 \caption{\fontsize{9}{9}\selectfont The joint 95\% posterior probability contours for the DM benchmarks 5 (green), 5a (red) and 5b (yellow) using data III and marginalising over the astrophysical uncertainties. For clarity the contours corresponding to benchmarks 5a and 5b were omitted in the top left frame. The only two parameters that can be well constrained within their priors are the DM mass $m_\chi$ and the inelastic parameter $\delta$.}\label{figall}
\end{figure}

 In order to make the analysis tractable, we have made thus far some simplifying assumptions, namely $\sigma_p^{SD}=\sigma_n^{SD}=0$ in Section \ref{secSI} and $f_n/f_p=1$ in Sections \ref{secSISD} and \ref{secSId}. Here, we let all parameters in Table \ref{tabPars} (except for $\sigma_n^{vec}$ and $b_p/b_n$) vary and identify which are the ones that can be robustly pinpointed. Adopting benchmark 5 as the true model and using $^{19}$F+$^{23}$Na+$^{40}$Ar+$^{73}$Ge+$^{127}$I+$^{131}$Xe (data III), we scan over the entire parameter space (check Table \ref{tabPars}) and marginalise over astrophysical uncertainties obtaining the results displayed in Figure \ref{figall}. It turns out that only two WIMP properties can be measured with good accuracy: $m_\chi$ and $\delta$. The respective $2\sigma$ ranges read $m_\chi=52\pm 20$ GeV and $\delta=35\pm 15$ keV. These figures are very robust against astrophysical uncertainties and rely on a minimum of theoretical assumptions. Apart from $m_\chi$ and $\delta$, all other parameters are left essentially unconstrained within their priors. For completeness we show in the top right and bottom plots of Figure \ref{figall} the cases of benchmarks 5a ($m_\chi=25$ GeV) and 5b ($m_\chi=250$ GeV). While for 5a both $m_\chi$ and $\delta$ can be reasonably well pinpointed, for 5b only the mass splitting $\delta$ can be measured. It is interesting to notice that for $m_\chi=250$ GeV the SD cross-section $\sigma_p^{SD}$ is systematically underestimated.

\par Finally, we repeat the scan for benchmark 5 but this time with a ``flat'' astrophysical setup (as defined in Section \ref{secmeth}). This is an useful procedure to check the dependence of WIMP signals on astrophysical parameters. We find that $m_\chi$ and $\delta$ can still be robustly determined, although with larger uncertainties than stated in the previous paragraph. Unfortunately, however, in this case a multi-target detection with the next generation of ton-scale instruments is not enough to constrain the astrophysical parameters $\rho_0$, $v_0$, $v_{esc}$ and $k$ -- in fact, the posterior distribution on these quantities is far wider than the priors used for the astrophysical uncertainty in Table \ref{tabPars}.

\section{Conclusion}\label{secconc}

\par In this work we have fully explored the ability of the next generation of ton-scale direct detection experiments to constrain WIMP properties. Using realistic upcoming experimental capabilities and including astrophysical uncertainties, we have studied how target complementarity can be used to improve WIMP-related constraints in case of a multiple-target analysis. Our main findings -- outlined in Table \ref{tabconc} --  may be summarised as follows:
\begin{itemize}
\item It will be possible to determine the sign of the ratio of scalar couplings $f_n/f_p$, but not its absolute value. This entails an uncertainty on the scalar-proton cross-section $\sigma_p^{sc}$ of more than one order of magnitude ($8\times10^{-11}\textrm{ pb}\lesssim \sigma_p^{sc} \lesssim 6\times10^{-9}\textrm{ pb}$), if we are to relax the widely used (but seldom tested) assumption $f_n/f_p=1$.

\item Scalar and vector cross-sections cannot be isolated even using multi-target data. Thus, the prospects for discriminating Majorana and Dirac WIMPs with upcoming ton-scale instruments are rather pessimistic.

\item The absolute value of $a_n/a_p$ can be fairly well constrained. However, we found that even with ton-scale direct detection experiments, two disconnected regions in the parameter space $(a_n/a_p,\sigma_p^{SD})$ are compatible with the mock data: one at $a_n/a_p<0$ (the ``correct'' solution) and one at $a_n/a_p>0$ (the ``wrong'' solution). The latter can be entirely discarded making use of targets such as $^{23}$Na and $^{127}$I which are particularly complementary to $^{19}$F, $^{73}$Ge and $^{131}$Xe. But we could only isolate the solution at $a_n/a_p<0$ with exposures for $^{23}$Na and $^{127}$I of 3 ton.yr that are probably out of reach within the next decade.

\item The inelastic parameter $\delta$ can be measured to an accuracy of tens of keV. This statement holds even for DM masses as large as 250 GeV in which case the mass itself cannot be precisely measured. Furthermore, inelastic DM candidates with modest mass splittings of $\delta\simeq 40$ keV can be easily distinguished from the standard elastic scattering scenario.

\item The only WIMP parameters that can be extracted from the data in a very robust manner are the DM mass $m_\chi$ and the inelastic parameter $\delta$. All other parameters can only be constrained at the expenses of introducing theoretically motivated hypotheses regarding WIMP couplings.
\end{itemize}

\begin{table*}[tp]
\centering
\fontsize{11}{11}\selectfont
\begin{tabular}{ccc||cccccc}
\hline
\hline
Section & data set & benchmark & $m_\chi$ [GeV] & $f_n/f_p$ & $\sigma_p^{sc}$ [$10^{-9}$ pb] & $a_n/a_p$ & $\sigma_p^{SD}$ [$10^{-5}$ pb] & $\delta$ [keV] \\
\hline
V.A & I & 1 & $49\pm11$ & $1.7\pm3.0$ & $0.5_{-0.43}^{+3.05}$ & -- & -- & -- \\

V.B & IV & 3 & $49\pm7$  & -- & $1.0_{-0.41}^{+0.70}$  & $-0.97\pm0.40$ & $1.0_{-0.44}^{+0.78}$ & -- \\

V.C & I & 4 & $56\pm30$ & -- & $1.0_{-0.68}^{+2.09}$ & -- & -- & $36\pm17$ \\

V.D & III & 5 & $52\pm20$ & $0.8\pm 4.8$ & $0.5_{-0.45}^{+4.29}$ & $0.01\pm 2.62$ & $0.13_{-0.12}^{+2.39}$ & $35\pm15$ \\

\hline
\end{tabular}
\caption{\fontsize{9}{9}\selectfont The $2\sigma$ predicted measurements for the next generation of ton-scale direct detection experiments assuming different DM benchmarks and data sets. In all cases presented here, astrophysical uncertainties are marginalised over according to Table \ref{tabPars}.}\label{tabconc}
\end{table*}

\par To conclude, in view of the next generation of ton-scale instruments, it is increasingly important to identify what direct detection can really tell us about WIMPs independently from any theoretical prejudice (regarding couplings, for instance) and keeping account of astrophysical uncertainties. Only that way may one hope to convincingly pinpoint the nature of dark matter.

\vspace{0.5cm}
\par {\it Acknowledgements:} It is a pleasure to thank Gianfranco Bertone for the continuous motivation, useful suggestions and proof-reading the manuscript, as well as Louis Strigari and Roberto Trotta for kindly providing private access to their direct detection code, numerous fruitful discussions and the careful reading of the paper. The author also acknowledges Chiara Arina, Nicolao Fornengo and Annika Peter for helpful comments and suggestions, and thanks the anonymous Referee for several suggestions and remarks that helped improving the article. The early stages of this work were supported by Funda\c{c}\~{a}o para a Ci\^encia e Tecnologia (Portuguese Ministry of Science, Technology and Higher Education) under the program POPH co-financed by the European Social Fund, and the later ones by the Swiss National Science Foundation.

\bibliographystyle{hunsrt}%organise refs for the first time referenced, with eprint number
\bibliography{tonscale_v3}

\end{document}